\documentclass[%
 reprint,
floatfix,
superscriptaddress,
shofootinbib,
 amsmath,amssymb,
 aps,
pra,
]{revtex4-1}
\DeclareUnicodeCharacter{2212}{-}
\usepackage{xcolor}
\usepackage{times}
\usepackage{tikz}
\usepackage{amssymb}
\usepackage{comment}
\usepackage{dsfont}
\usepackage{graphicx}
\usepackage{dcolumn}
\usepackage{bm}
\usepackage{epstopdf}
\usepackage[colorlinks]{hyperref}
\hypersetup{colorlinks = true,
linkcolor = blue,
filecolor = blue,
urlcolor = blue,
citecolor = blue}

\begin{document}
\title{Impurity-induced Mott ring states and Mott zeros ring states in the Hubbard operator formalism}
\author{Emile Pangburn}
\affiliation{Universit\'e Paris Saclay, Institut de Physique Th\'eorique, CEA CNRS,
Orme des Merisiers, 91190 Gif-sur-Yvette Cedex, France}
\author{Anurag Banerjee}
\affiliation{Universit\'e Paris Saclay, Institut de Physique Th\'eorique, CEA CNRS,
Orme des Merisiers, 91190 Gif-sur-Yvette Cedex, France}
\author{Catherine P\'epin}
\affiliation{Universit\'e Paris Saclay, Institut de Physique Th\'eorique, CEA CNRS,
Orme des Merisiers, 91190 Gif-sur-Yvette Cedex, France}
\author{Cristina Bena}
\affiliation{Universit\'e Paris Saclay, Institut de Physique Th\'eorique, CEA CNRS,
Orme des Merisiers, 91190 Gif-sur-Yvette Cedex, France}

\begin{abstract}
We study the formation of subgap impurity states in strongly correlated Mott insulators. We use a composite operator method that gives us access to both the bulk Green's function, as well as to the real-space Green's function in the presence of an impurity. Similar to the non-interacting systems, we show that the formation of impurity subgap states at large impurity potential (``Mott ring states") depends rather on the band-mixing, than on the topological character of the system. Thus even a trivial Mott insulator can under certain conditions exhibit ring states. For the system studied here the band mixing is that between the holon and doublon elementary excitations rather than an orbital mixing. Moreover we study the formation of bands of zeros in the correlated Green's function, believed to exhibit a free quasiparticle-like behavior. We show that in the presence of an impurity the same conclusion can be applied, i.e. ``Mott zeros ring states" form in the presence of topological bands of zeros, but also for trivial quasi-flat bands of zeros with band mixing.
\end{abstract}

\maketitle
\section{Introduction\label{introduction}}
Topological phases represent a class of matter that cannot be described by the traditional Landau paradigm of local order parameters~\cite{AltlandZirnbauer,bernevig_book}. A key implication of topological phases is the bulk-boundary correspondence, which guarantees the existence of gapless edge states at the interface of regions with differing topological invariants~\cite{bernevig_book,KaneMele}. A broader framework, known as bulk-defect correspondence~\cite{teo2010topological, teo2017topological,jurivcic2012universal,benalcazar2019quantization,li2020fractional,geier2021bulk,liu2021bulk,roy2021dislocation}, connects topological classes to protected zero-energy modes associated with topological defects. While local potential impurities are not classified as topological defects, a connection remains between impurity-induced subgap states for strong impurities and non-trivial topology~\cite{liu2009classical,shan2011vacancy,wang2012impurity,sau2013bound,slager2015impurity,mashkoori2017impurity,mashkoori2019impact,diop2020impurity,michel2023bound}. 
Recent studies have shown that these so-called "ring states"~\cite{queiroz2024ring} arise due to orbital band mixing, a fundamental requirement for the formation of topological bands. This connection explain the emergence of ring states although it does not imply an equivalence between topology and ring states. These considerations are essential for making concrete predictions that enable the distinction between trivial and topological insulators using real-space spectroscopy.


We are interested here to address this question in topological strongly correlated systems. 
Numerous studies have examined the transition of trivial Mott excitations to anomalous quantum Hall insulators with generic Hubbard interactions~\cite{YoshidaBHZ,Hohenadler_2013,KMHubb0,KMHubb2,LeeBHZ,he2024phase}. Recently, Moir\'e materials have also emerged as a promising platform for realizing topological bands in strongly correlated electron systems~\cite{Li2021,SerlinExp,Chen2020}. In correlated systems, topology is encoded in the ground state properties, but extracting it is often computationally challenging. In many cases, only the excitation spectrum, as captured by the Green's function, can be computed. This poses a challenge in understanding how Green's functions encode the topology of strongly correlated models~\cite{he2016topological1,he2016topological2,zhao2023failure}. 

In Mott insulators, the poles of the two-point Green's function correspond to quasiparticle excitations. In a recent work~\cite{pangburn2024topological} we have explored such excitations using a composite operator framework~\cite{hubbard1963electron,Beenen0,avella1998hubbard,stanescu2000d,avella2011composite,Haurie_2024,pangburn2024spontaneous}. This formalism seeks to capture strong correlation effects by identifying a set of well-defined quasiparticles that characterize the fundamental excitations emerging from a non-trivial ground state. This quasiparticle set is composed of the holon ($\xi$) and doublon ($\eta$) operators. With only intra-orbital Coulomb interaction, and without aiming to determine the exact hopping-induced phase diagram, this quasiparticle set suffices to capture the simplest topological features of a Mott insulator, as demonstrated in Ref.~\cite{pangburn2024topological}.

In this work we investigate the role of local impurities as probes of topology in strongly correlated Mott insulators. 
We use a real-space framework based on Hubbard operators~\cite{banerjee2024charge,pangburn2024topological}. We find that the conclusions of the non-interacting analysis can be carried to Mott insulators, i.e. "Mott ring states" arise also mainly due to band mixing. This is no longer an orbital mixing, but stems rather from the hybridization of holons and doublons. The formation of Mott ring states will thus be almost automatic in topological Mott insulators that by definition exhibit such mixing. However, this connection does not imply an equivalence between topology and ring states, and we show that, same as for the non-interacting case, a trivial Mott insulator may exhibit Mott ring states in the presence of almost flat bands with band mixing.

Additionally, Green's function zeros naturally appear at each momentum as the electronic spectral weight is split between the lower and upper Hubbard bands in Mott insulators~\cite{dzyaloshinskii2003some,seki2017topological}. The bands of Green function zeros in Mott insulators are currently receiving a lot of interest because they can exhibit a finite winding number and manifest an equivalent bulk-boundary correspondence~\cite{gurarie2011single,gurarie2013topological,wagner2023edge,pangburn2024topological}. 
The zeros form bands that possess orbital and spin character, allowing for a classification in direct correspondence with the bands of non-interacting insulators, as described through space group representation theory~\cite{bradlyn2017topological,kruthoff2017topological}. Many studies have focused on translation-invariant systems in $k$-space
~\cite{hatsugai1992exactly,setty2024electronic,bollmann2024topological}, some
extending earlier investigations also towards orthogonal metals~\cite{nandkishore2012orthogonal}. Additionally, more quantitative analyses have been conducted using DMFT-like methods~\cite{stepanov2024interconnected}. 
 
While the topology of zero-energy bands has been well-characterized and classified using band representations~\cite{setty2024symmetry,chen2024dirac}, the response of these zeros to local probes remains an open question.
Since Green's function zeros obey a bulk-boundary correspondence and have a quasiparticle-like interpretation, a natural question arises: what is the nature of the bulk-impurity correspondence for Green's function zeros? We show that the Green's function bands of zeros behave also similarly to non-interacting bands, in that they give rise to ring states (``Mott zeros ring states'') in the presence of band mixing. These states arise either for bands of zeros with topological character, as well as for non-topological bands with an almost flat dispersion and band-mixing.


The models and the numerical methods used for the calculations are presented in Sec.~\ref{sec:Model_Method}. A review of the non-interacting results and our numerical findings for strongly interacting Mott insulators are presented in Sec.~\ref{sec:Res}. Finally, in Sec.~\ref{sec:Discussions}, we summarize our results and present some possible extensions of our work. Various technical details pertaining to our calculations are presented in the Appendix.
 
\section{Model and method\label{sec:Model_Method}}

We investigate the response of a single potential impurity as a probe of topological signatures in both non-interacting and strongly interacting models. The introduction of the impurity breaks the translation symmetry of the system. Here, we discuss the models and methods used to study the systems with a single potential impurity. 

In the Hubbard operator formalism employed to study the Mott insulating state, the two-point Green's function is the central object of the approximation. In the following, we distinguish non-interacting band insulators and Mott insulators by denoting the two-point Green's function as $\mathcal{G}$ in the interacting case and $\mathcal{G}_0$ in the non-interacting case.

\subsection{Model}
\subsubsection{Non-interacting models}
First, we study the Chern model on a square lattice as the simplest non-interacting system with topologically non-trivial bands. Historically, Haldane~\cite{Haldane} introduced the first Chern insulator on the honeycomb lattice. However, the Chern insulator model on a square lattice is closely related to the Haldane model and is straightforward to introduce. This is given by
\begin{align}
\mathcal{H}^{\rm C}_0=\sum\limits_{k} & \psi^\dagger \left[\mathbf{\Gamma} . \left(\hat{\tau} \otimes \hat{\sigma}_0 \right) - \mu \left(\hat{\tau}_0 \otimes  \hat{\sigma}_0\right) \right]  \psi \\
&\Gamma_x(k)=t\sin(k_x)&\\
&\Gamma_y(k)=t\sin(k_y)&\\
&\Gamma_z(k)=M+ t \left(\cos(k_x)+\cos(k_y) \right)&
\end{align}
Here $\psi$ is given by $\psi=\left(c_{ks\uparrow},c_{kp\uparrow}, c_{ks\downarrow},c_{kp\downarrow}\right)^T$ where $c^\dagger_{k\alpha\sigma}$ and $c_{k\alpha\sigma}$ are the creation and annihilation operators for an electron with crystal momentum $k$ and  orbital character $\alpha \in \{s,p\}$ and spin $\sigma \in \{\uparrow,\downarrow\}$. The operator $\hat{\tau}$ represent the usual Pauli matrices in the orbital basis, and $\hat{\sigma}$ are the same in the spin basis. The chemical potential $\mu$  fixes the average electron density to half-filling. 

The Hamiltonian explicitly breaks the time-reversal symmetry and captures integer quantum Hall physics without any external magnetic field. The topology of this Hamiltonian class is characterized by the Chern number $\mathcal{C}$ and such Hamiltonians belong to class A of the ten-fold classification of topological insulators. 

In what follows, it will also be useful to study a model with a trivial Chern number but a nontrivial topology protected by specific symmetries. We select a model from class $AII$ of the tenfold way, which gives rise to quantum spin Hall insulators with topology protected by time-reversal symmetry $\mathcal{T}$. The topological invariant for this symmetry class is known as the spin Chern number, denoted by $\mathcal{C}_S$. In the absence of spin-orbit coupling, it can be computed as $\mathcal{C}_S=\left(\mathcal{C}_\uparrow-\mathcal{C}_\downarrow \right)/2 $, where $\mathcal{C}_\sigma$ represents the Chern number for spin $\sigma$. However, when $U(1)$ spin-symmetry is broken, the topological class is preserved, but the calculation of the $\mathds{Z}2$ invariant becomes more intricate~\cite{prodan2009robustness, lin2024spin}. These models are constructed as the direct sum of two time-reversed copies of the Chern model, effectively restoring time-reversal symmetry by incorporating both spin-up and spin-down states.
As a specific example, we investigate the quantum spin Hall model on the square lattice, known as the Bernevig-Hughes-Zhang (BHZ) model~\cite{bernevig2006quantum}. This model is conceptually similar to the Kane-Mele model~\cite{KaneMele} on the honeycomb lattice and is explicitly given by:


\begin{align}
&\mathcal{H}_0^{\rm BHZ}= -\sum\limits_k\psi^\dagger \left[\mu \left(\hat{\tau}_0 \otimes  \hat{\sigma}_0\right) +\Delta_{\tau} (\hat{\tau}_x \otimes \hat{\sigma}_x) \right] \psi  & \nonumber \\
&+\sum\limits_{k}  \psi^\dagger \left[\mathbf{\Gamma}_x . \left(\hat{\tau}_x \otimes  \hat{\sigma}_z \right) +\mathbf{\Gamma}_y . \left(\hat{\tau}_y \otimes  \hat{\sigma}_0 \right)+\mathbf{\Gamma}_z . \left(\hat{\tau}_z \otimes  \hat{\sigma}_0 \right)  \right]\psi \nonumber &\\
&+\alpha^{SOC}\sum\limits_k \psi^\dagger\left[\cos(k_x)+\cos(k_x)\right]\left( \hat{\tau}_0\otimes \hat{\sigma}_x\right)\psi&
\label{eq:BHZ_model}
\end{align}

The standard BHZ model corresponds to $\Delta_\tau = 0$ and $\alpha^{SOC}=0$, which ensures that the Hamiltonian preserves time-reversal symmetry and $U(1)$ spin symmetry. The topology in this case is described by the $\mathds{Z}2$ invariant~\cite{KaneMele}. However, when $\Delta_\tau \neq 0$, time-reversal symmetry is broken, and the topology of the model is then characterized by the Chern number $\mathcal{C}$. When $\alpha^{SOC}\ne 0$, time-reversal symmetry is preserved but $U(1)$ spin symmetry is broken. We extend the Green's function method introduced in Ref.~\cite{setty2024electronic} to compute the $\mathds{Z}2$ invariant associated with the single-particle Green's function. A detailed description of this approach can be found in App.\label{App:G_Spin_Topo}.  \\

\subsubsection{Strongly correlated models}
In this work, we investigate also strongly correlated models that exhibit Mott insulating behavior at half-filling. To study this phase, we incorporate strong on-site intra-orbital Hubbard interactions into the non-interacting band models described above. The interacting term is written as
\begin{align}
\mathcal{H}_{U}=\sum\limits_\mathbf{i,\alpha} U_\alpha \hat{n}_{i \alpha \uparrow} \hat{n}_{i\alpha\downarrow}
\end{align}
The parameters $U_\alpha$ are the on-site repulsion on the $s$- and $p$-orbitals, which can be different in general.  The ${\hat{n}_{i\alpha\sigma}=c^\dagger_{i\alpha\sigma}c_{i\alpha\sigma}}$ is the number operator. \\

\subsubsection{Impurity}
The main focus of this work is the response of a given system to localized impurities. We introduce an impurity potential at a specific lattice site $i_0$ by adding the following term to the Hamiltonian:
\begin{align}
    \mathcal{H}_V=  V_{\alpha\sigma} c^\dagger_{i_0\alpha\sigma}c_{i_0\alpha\sigma},
\end{align}
here $V_{\alpha\sigma}$ represents the impurity strength for orbital $\alpha$ and spin $\sigma$. We consider two types of impurities: scalar impurity, the strength is identical for the two orbitals, and spins i.e., $V_{\alpha \sigma} = V_{\rm imp}$ for both orbitals and spin and magnetic impurities, such that $V_{\alpha \uparrow} = -V_{\alpha \downarrow}= V_{\rm imp}$ for both orbitals.

\subsection{$T$-matrix method for non-interacting systems\label{sec:T_matrix_review}}
In non-interacting finite systems, one can extract the impurity states numerically by exactly diagonalizing the Hamiltonian. However, for localized impurities and infinite systems, the problem can also be solved fully analytically  using the so-called $T$-matrix method~\cite{balatsky2006impurity}.
The central object of the method is the frequency-dependent $T-$matrix which is expressed as a function of the impurity potential matrix $\hat{V}$ (described by the elements $V_{\alpha\sigma}$),  and the momentum-dependent electronic Green function $\mathcal{G}_0(\mathbf{k},\omega)$
\begin{align}
&\mathcal{G}_0(\mathbf{k},\omega)=\left(\omega-\mathcal{H}_0(\mathbf{k})+i\epsilon\right)^{-1}&\\
&T(\omega)=\left(\mathds{1}-\hat{V}\int d\mathbf{k}\mathcal{G}_0(\mathbf{k},\omega)\right)^{-1}\hat{V}.&\label{eq:T_matrix}
\end{align}
Here $\mathcal{H}_0(\mathbf{k})$ is the tight-binding Hamiltonian of the non-interacting model, and $\epsilon$ is a damping parameter, equivalent to the quasi-particle inverse lifetime.

We study the average correction to the density of states $\delta\rho(\omega)$ induced by the impurity:
\begin{align}
&\delta\rho(\omega)=\text{Tr}\left(\int d\mathbf{k}\text{Im}\left[\mathcal{G}_0(\mathbf{k},\omega)T(\omega)\mathcal{G}_0(\mathbf{k},\omega)\right]\right).&
\end{align}
The spin-resolved density of states correction along the $S_z$-spin direction, which is useful for class $AII$ model, can also be extracted as
\begin{align}
&\delta\rho_\sigma(\omega)=\text{Tr}\left(\hat{\sigma}_z\int d\mathbf{k}\text{Im}\left[\mathcal{G}_0(\mathbf{k},\omega)T(\omega)\mathcal{G}_0(\mathbf{k},\omega)\right]\right)&
\end{align}
where $\hat{\sigma}_z$ is the spin projection operator along the $\hat{z}$-axis, with eigenvalues $+1$ for spin-up and $-1$ for spin-down. 

In the absence of impurities the density of states  $\rho_0( \omega)$ and the spectral function $A(\mathbf{k}, \omega)\propto -\text{Im }\rm{Tr}[ \mathcal{G}_0(\mathbf{k}, \omega) ]$ are both zero within the gap. Subgap impurity states correspond to a non-zero density of states within the gap. From Eq.~(\ref{eq:T_matrix}), subgap states can appear at very large impurity strength only if the following condition is satisfied:
\begin{align}
& \det \left[\int d\mathbf{k} \mathcal{G}_0(\mathbf{k}, \omega_0)\right] = 0,&
\end{align}
for at least one $\omega_0$ energy value inside the gap. This condition can be equivalently formulated as a requirement for the {\it local} Green's function $\det[\mathcal{G}_0(\mathbf{R}=\mathbf{0},\omega)=\int d\mathbf{k}\mathcal{G}_0(\mathbf{k},\omega)]$
to have zeros within the gap. Moreover, in the limit of very large impurity strength, any impurity-induced state with energy within the gap cannot have a finite weight at the impurity site, as its energy would diverge. Consequently, such impurity states develop a ring-like structure in real space, and thus can be referred to as "ring states"~\cite{queiroz2024ring}. 


\subsection{Mott insulator and composite operator method}

To investigate the ring states induced by local impurities in Mott insulating phases, we focus on systems where the intra-orbital on-site repulsion is significantly larger than the electronic bandwidth. An important point is that in this Mott insulating state, the electronic spectral weight is split across the lower and upper Hubbard bands~\cite{phillips2006mottness}. 
This fact leads to the possibility of the existence of $k$-space Green's function zeros, for which the interacting $\mathcal{G}(\mathbf{k},\omega)$ satisfies
\begin{align}
&\text{det}\left[\mathcal{G}(\omega,\mathbf{k})\right]=0,&
\end{align}
a feature that cannot occur in a non-interacting model. 


In order to study Mott insulators, we employ the composite operator method (COM)~\cite{avella2011composite}, which decomposes electron operators into a sum of two weakly interacting excitations: holons ($\xi$) and doublons ($\eta$). A detailed review of this method, including all technical aspects, can be found in Ref.~\cite{pangburn2024spontaneous,pangburn2024topological}. \\

The $T$-matrix method described in Sec.~\ref{sec:T_matrix_review} for non-interacting models is no longer applicable in strongly correlated phases. As a result,
the impurity calculations must be performed directly in real space. All technical details on handling the Hubbard operator method for inhomogeneous systems are provided in App.(\ref{App:Review_COM}). This real-space framework has been previously discussed in Ref.\cite{pangburn2024topological}. The calculations were performed on a $32 \times 32$ system for the Chern Mott insulator and $24 \times 24$ for the BHZ Mott insulator. 


For a Mott insulator, instead of using $\delta\rho(\omega)$, following the approach in~\cite{wagner2023edge}, we propose analyzing the eigenvectors of the electronic Green's function in the presence of an impurity, $\mathcal{G}_V(\omega,\mathbf{r}_i,\mathbf{r}_j)$ at each frequency $\omega$. By examining the spectrum $|\lambda(\mathcal{G}_V)|$ of $\mathcal{G}_V$, we identify its maximal and minimal eigenvalues, $\max\left(|\lambda(\mathcal{G}_V)|\right)$ and $\min\left(|\lambda(\mathcal{G}_V)|\right)$. A maximum eigenvalue corresponds to a Green's function pole, while a minimum represents a Green's function zero. 
In the case of poles, the density of states and $\max\left(|\lambda(\mathcal{G}_V)|\right)$ serve as equivalent probes. 


\section{Results}
\label{sec:Res}

\subsection{Ring state in non-interacting models\label{sec:Ring_NI}}
Before analyzing the impurity response in strongly correlated Mott insulators, we first review the mechanisms driving the formation of ring states in non-interacting insulators, as recently proposed in Ref.~\cite{queiroz2024ring}. 

In Appendix~\ref{App:Ring_Flat}, we derive an exact formula for the location of {\it local} Green's function zeros as a function of orbital mixing in the case of exactly flat bands. We remind the reader that the {\it local} Green's function zeros correspond to $\det[\mathcal{G}_0(\mathbf{R}=0]=0$, and are completely different from the Green's function bands of zeros that correspond to zeros of the Green's function in momentum space. A non-interacting system may exhibit {\it local} Green's function zeros but no Green's function bands of zeros. 

It appears that the presence of {\it local} Green's function zeros is controlled rather by the orbital mixing, then by the topological nature of the system. Thus, in this extreme flat band case, the tiniest amount of orbital mixing will lead to {\it local} Green's function zeros within the gap, which proves that non-topological states can also possess ring states for large impurity strengths. Since topological phases necessitate orbital mixing, as exemplified by the non-trivial Chern number, the presence of wavefunction zeros~\cite{kohmoto1985topological} and of {\it local}  Green's function zeros within the gap is very likely for topological insulators. 

Note that, however, even topological insulators are not guaranteed to have {\it local}  Green's function zeros within the single-particle gap. The position of these Green's function zeros correspond to those of the diagonal components of the Green's function matrix and thus to the center of mass of the electronic weight for a given orbital character. This is true provided that {\it local}  off-diagonal terms vanish, such as when the orbitals are orthogonal. Moreover, similar to probes like superfluid weight~\cite{liang2017band}, there are two main contributions: a geometric contribution, which arises solely from the eigenvectors, and a conventional contribution, which also takes into account the band dispersion. The conventional contribution is acting against the geometric one, shifting the position of the {\it local} Green's function zeros closer to the bands. In the case of flat bands, only the geometric contribution remains, making the observation of {\it local}  Green's function zeros within the gap easier. However, one can conceive a situation in which the bandwidth contribution may overcome the geometric one and thus even a topological insulator may end up with no Green's function zeros within the gap. However such situation is very unlikely and there is no known example at this point.

\begin{figure}[h!]
\includegraphics[width=9cm]{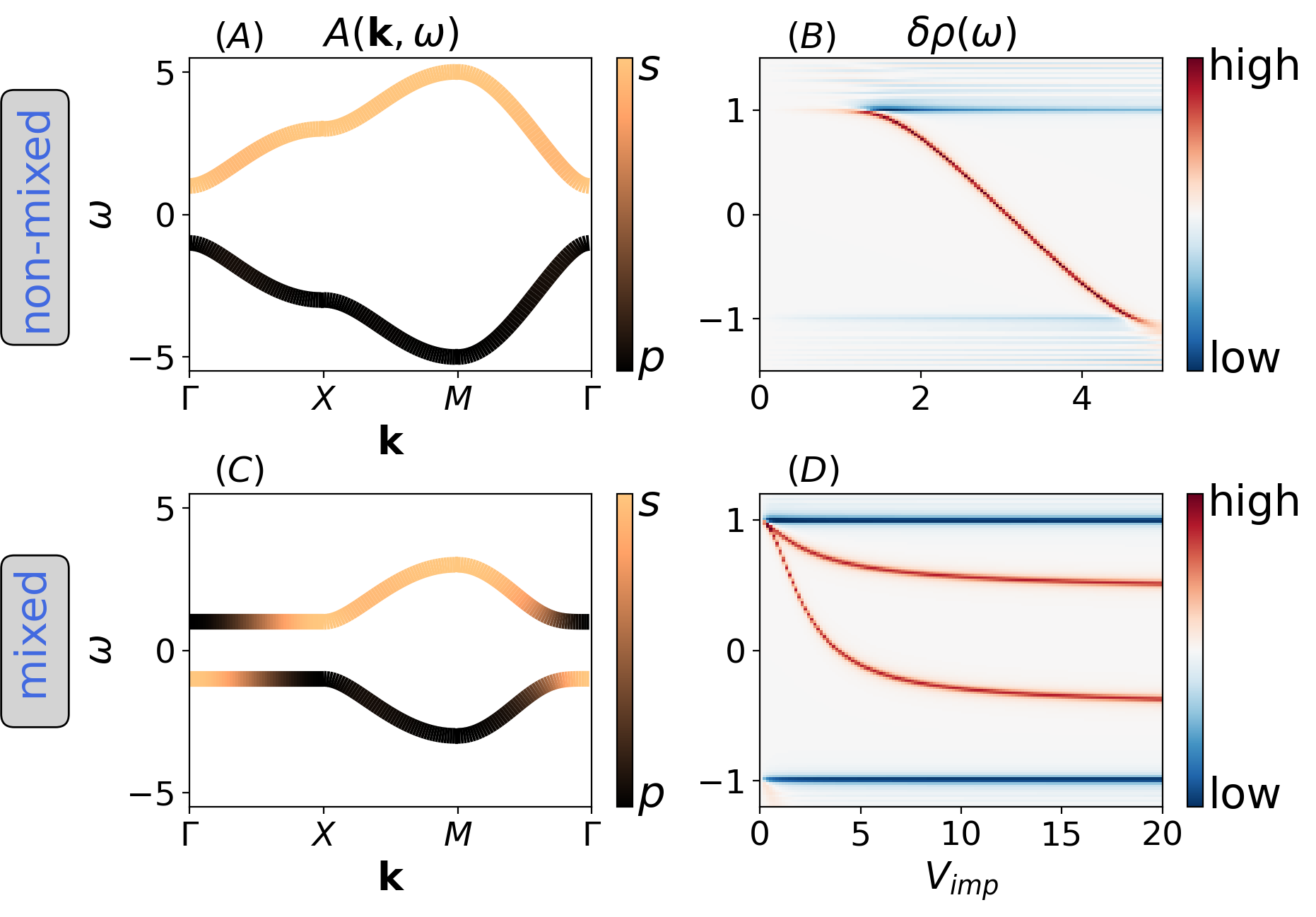}
\caption{Formation of impurity-induced subgap states in a two-band model with finite bandwidth. Panels (A), and (C) illustrate orbitally resolved bulk bands for a two-band Chern insulator: (A) trivial bands with uniform orbital character ($M=3.0t$), and (C)  topological bands ($M=1.0 t$). In panels (B), and (D), the corresponding correction of the average density of states $\delta\rho(\omega)$ is shown as a function of impurity strength. For this example only the topological phase exhibits stable subgap states.}
\label{fig:fig1}
\end{figure}

We illustrate the formation of the ring states in Fig.~\ref{fig:fig1}. In the upper row, we present the orbital character of the bands and $\delta\rho(\omega)$ for a trivial band insulator, whereas the lower row illustrates the corresponding features for a topological insulator. We observe that when the bands exhibit a uniform orbital character, no ring states appear within the energy gap. In contrast, in the topological case for which the bands have a mixed orbital character, ring states emerge.

In the following, we extend this principle to strongly correlated Mott insulators.

\subsection{Strongly correlated Mott insulator\label{sec:Ring_Mott}}
The Hubbard operator framework requires to be supplemented by an approximate decoupling scheme.  In this work, along the lines of our previous work on topological Mott insulators~\cite{pangburn2024topological}, we utilize the Roth decoupling~\cite{hubbard1963electron, hubbard1964electron, roth1969electron, avella1998hubbard}, which is outlined in App~\ref{App:Review_COM}. This leads to self-consistent equations that must be solved in real space when translational symmetry is broken by the introduction of an impurity. Self-consistent solution or single-shot computation starting from the translation invariant solution are giving very close results. This implies that the self-consistent framework is not essential for understanding the physics of ring states in Mott insulators, provided that the single-particle gap remains sufficiently large and does not locally close at high impurity strength.
In the following section, we provide single-shot computation to illustrate that the underlying physics relies solely on two key assumptions: first, that the electron is no longer an elementary excitation in a Mott insulator, and second, that the considered excitations, the holon $\xi$ and the doublon $\eta$, are well-defined quasiparticles. This conclusion is independent of the specific technical details of the composite operator method (COM).

\begin{figure}[h!]
\includegraphics[width=7cm]{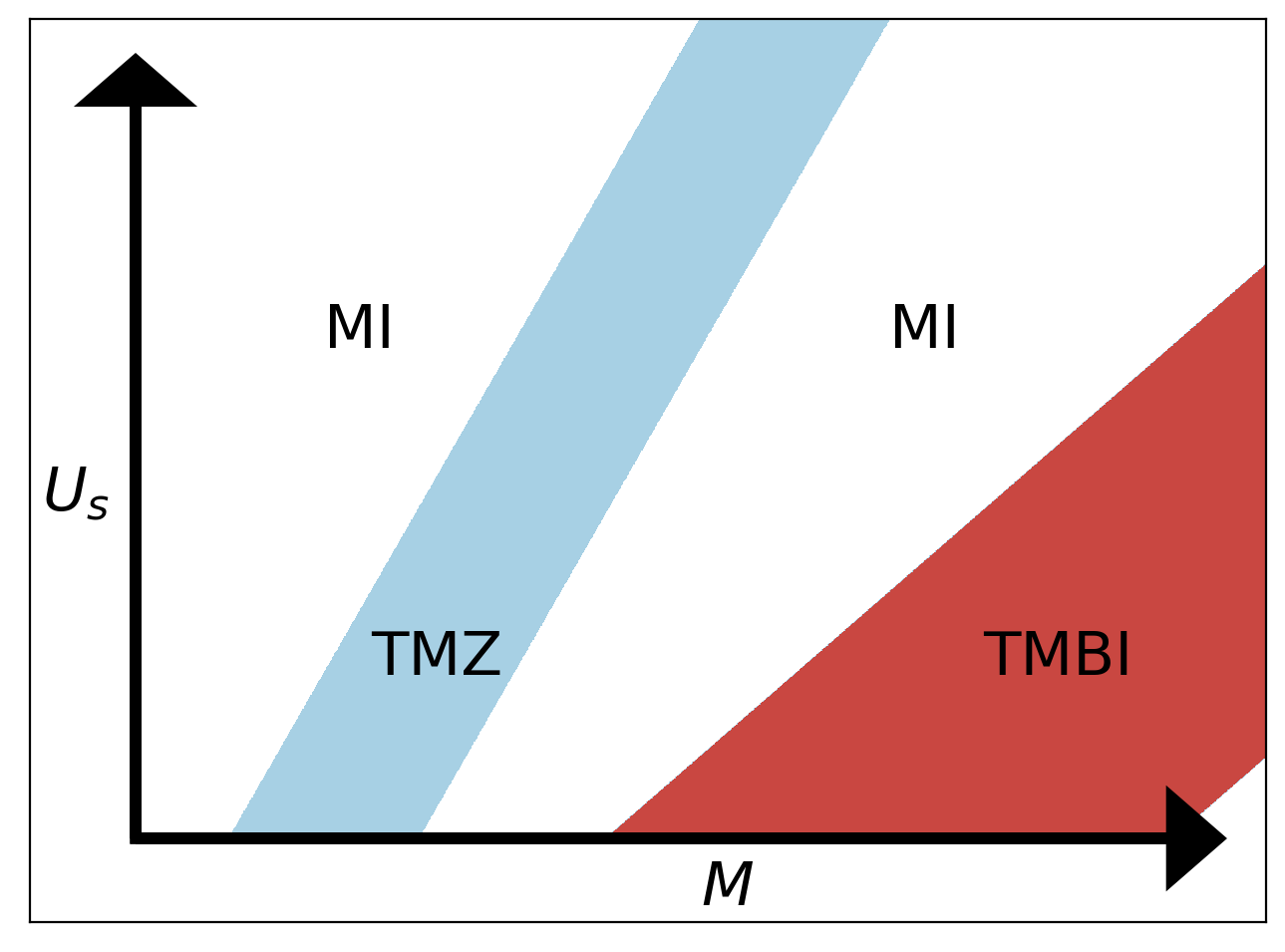}
\caption{Schematic representation of the self-consistent phase diagram from Ref.~\cite{pangburn2024topological}, showing the dependence on the crystal field $M$ and the intra-orbital Coulomb repulsion $U_s$, while keeping the Coulomb repulsion on the other orbital $U_p$ fixed. The white region of the phase diagram represents a trivial Mott insulator (MI). The blue region corresponds to the TMZ phase, characterized by topological Green's function zeros, while the red region denotes the TMBI phase, where the topology is associated with Green's function poles.}
\label{fig:fig1prime}
\end{figure}

As mentioned earlier, the Hubbard operator method is a Green's function approach based on the equation of motion technique. Consequently, topological properties must be extracted from the single-particle Green's function. For $2d$ tight-binding models, the winding number of the Green's function, denoted as $N_3[\mathcal{G}]$, can be computed to characterize the system's topology. The technical details are presented in App.~\ref{App:G_Topo_Review}. A phase is then called topological when $N_3[\mathcal{G}]\ne 0$. Because in a Mott insulator, single-particle Green's function are possessing both poles and zeros, the non-triviality of $N_3[\mathcal{G}]$ can either comes from zeros or poles. Within self-consistent calculations, no phase featuring both topological poles and topological zeros could be stabilized~\cite{pangburn2024topological}, as illustrated schematically in Fig.~\ref{fig:fig1prime}. When $N_3[\mathcal{G}]=0$, the system is classified as a Trivial Mott Insulator (MI). If $N_3[\mathcal{G}] \ne 0$ due to contributions from poles, we refer to it as a Topological Mott Band Insulator (TMBI). Conversely, if $N_3[\mathcal{G}] \ne 0$ due to contributions from zeros, the phase is termed a Topological Mott Zeros (TMZ) phase. 

\begin{figure*}[h!]
\includegraphics[width=18cm]{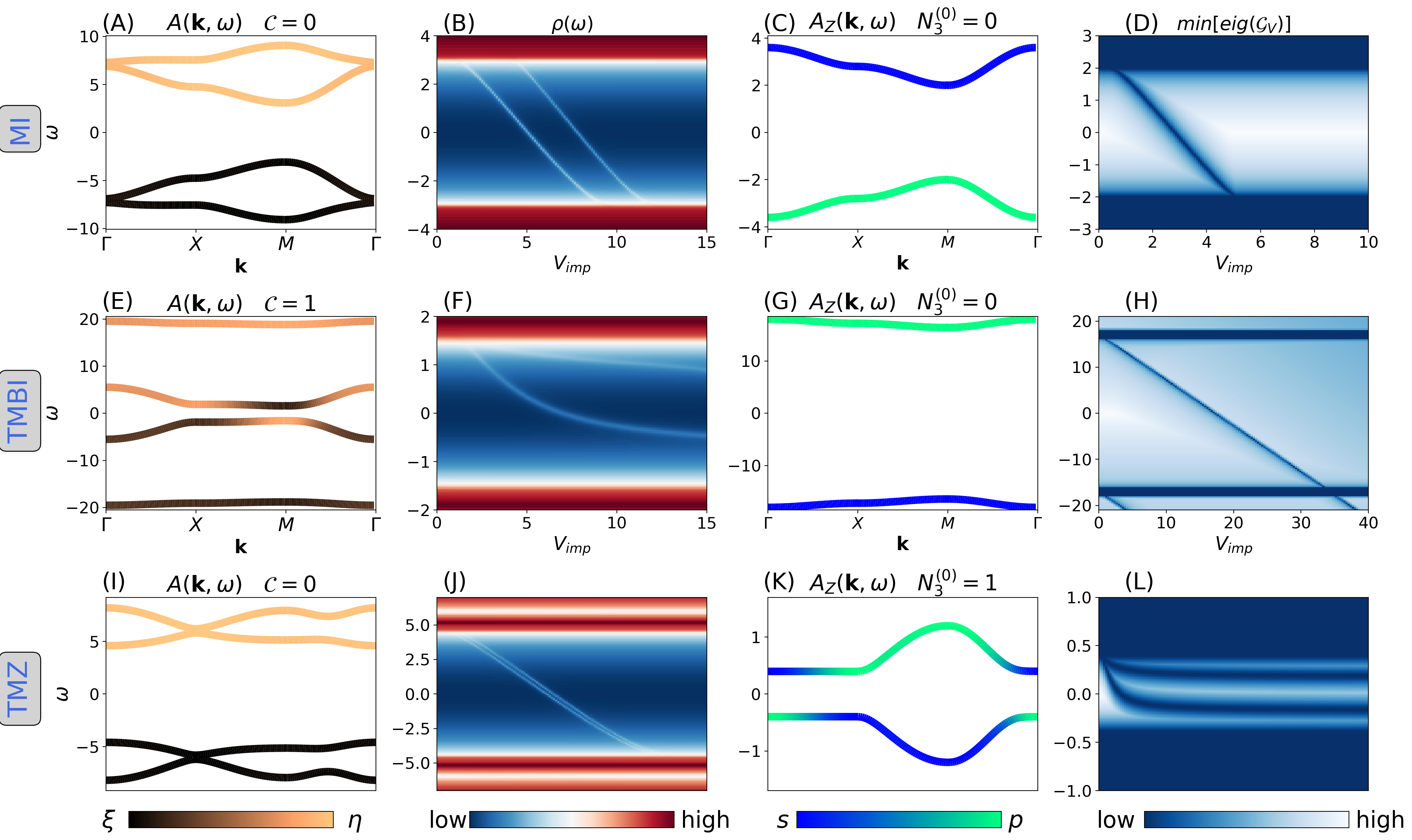}
\caption{Formation of impurity-induced subgap states in the Chern-Hubbard insulator. The first column depicts the orbitally resolved bulk bands (also denoted bands of poles). The third column depicts the orbitally resolved bands of Green's function zeros (also denoted bands of zeros). 
The first line corresponds to a trivial Mott insulator, characterized by trivial bands of both poles and zeros. The second line illustrates the TMBI phase, where the poles exhibit topological behavior and a mixed orbital character, while the zeros remain trivial. The third line corresponds to the TMZ phase, for which the poles are topologically trivial, while the zeros are topological and exhibit a mixed orbital character. 
The panels in the second column depict the correction to the average density of states, $\delta\rho(\omega)$, as a function of energy and impurity strength. Each panel corresponds to a band-of-poles configuration from the first column. Similarly the panels in the fourth column depict $\min[\rm{eig}(\mathcal{G}_V)]$ as a function of energy and impurity strength, each panel corresponding to a configuration for a band of zeroes from the third column. The main observation is the formation of Mott ring states and correspondingly of Mott zeros ring states in the presence of topological sets of bands with band mixing.  The COM parameters used to generate this figure are given in App.~\ref{App:Params}.}
\label{fig:fig2}
\end{figure*}

In the following, we first analyze the local impurity response for each of these three distinct phases within the Chern-Hubbard model. 
In Fig.~\ref{fig:fig2}, we present a summary of the impurity response results for a representative set of parameters corresponding to each distinct topological class for the Chern-Hubbard model. In each case, we showcase the orbitally resolved edge state structure for both poles and zeros, highlighting the mixed orbital character of the bands and the topological character of the phase. Additionally, we display the density of states, $\rho(\omega)$, for poles, along with the minimum eigenvalues $\min\left(|\lambda(\mathcal{G}_V|\right)$ for zeros.

In the MI phase shown in the upper row of Fig.~\ref{fig:fig2}, where both poles and zeros possess a trivial topological invariant, orbital mixing remains weak. As a result, no ring states appear at large impurity strengths for energies within the gap in either the poles or zeros spectrum. The only notable difference from the non-interacting case is the doubling of the number of impurity-induced pole states, which directly arises from the assumption that electrons split into holons and doublons, effectively doubling the number of impurity states. Because Green's function zeros can be interpreted as quasiparticles formed by bound holon-doublon states~\cite{pangburn2024topological}, there is no doubling of the zero states. These zeros correspond to bound states with both $s$- and $p$-orbital character, leading to results that are identical to those of a trivial non-interacting insulator.

In the TMBI phase corresponding to the middle row, band inversion gives rise to topological poles as shown in Fig.~\ref{fig:fig2}.E. An important point to note is that the band inversion involves only two bands, while the other lower and higher energy bands remain orbitally uniform. In this case, the low-energy physics is similar to that of a non-interacting topological insulator, which is reflected in $\rho(\omega)$ where we observe the appearance of ring states inside the energy gap. The key difference lies in the interpretation of these states, as they possess a mixed holon-doublon character at the origin of the band inversion. Concerning the zeros in this phase, they exhibit a uniform orbital character, which results in the absence of subgap zero states at large impurity strengths. 

Finally, the last phase we examine is the TMZ phase corresponding to the bottom row. As shown in Fig.~\ref{fig:fig2}, the poles have uniform orbital character in this phase. Similar to the MI phase, due to the doubling of electrons into doublons and holons, we observe two impurity states, which, at large impurity strength, do not remain inside the gap. The topological nature of the zeros is reflected in the orbitally mixed bands as shown in Fig.~\ref{fig:fig2}.K. Similar to the MI and TMBI phases, the zeros in this phase follow the behavior of a non-interacting band insulator. Specifically, we observe two zero-ring states at large impurity strength, akin to what would be seen in a topological 2-band Chern insulator. 

The main conclusion that can be drawn is that the ring state principle, reviewed in Sec.~\ref{sec:Ring_NI} for the non-interacting model, also applies to Mott insulators described through holons and doublons. This applies to both Green's function poles and zeros, which can be interpreted as the dispersion of emergent quasiparticles, bound states of holons and doublons. 

\subsection{Ring states in a topologically trivial Mott insulator}

\begin{figure}[h!]
\includegraphics[width=9cm]{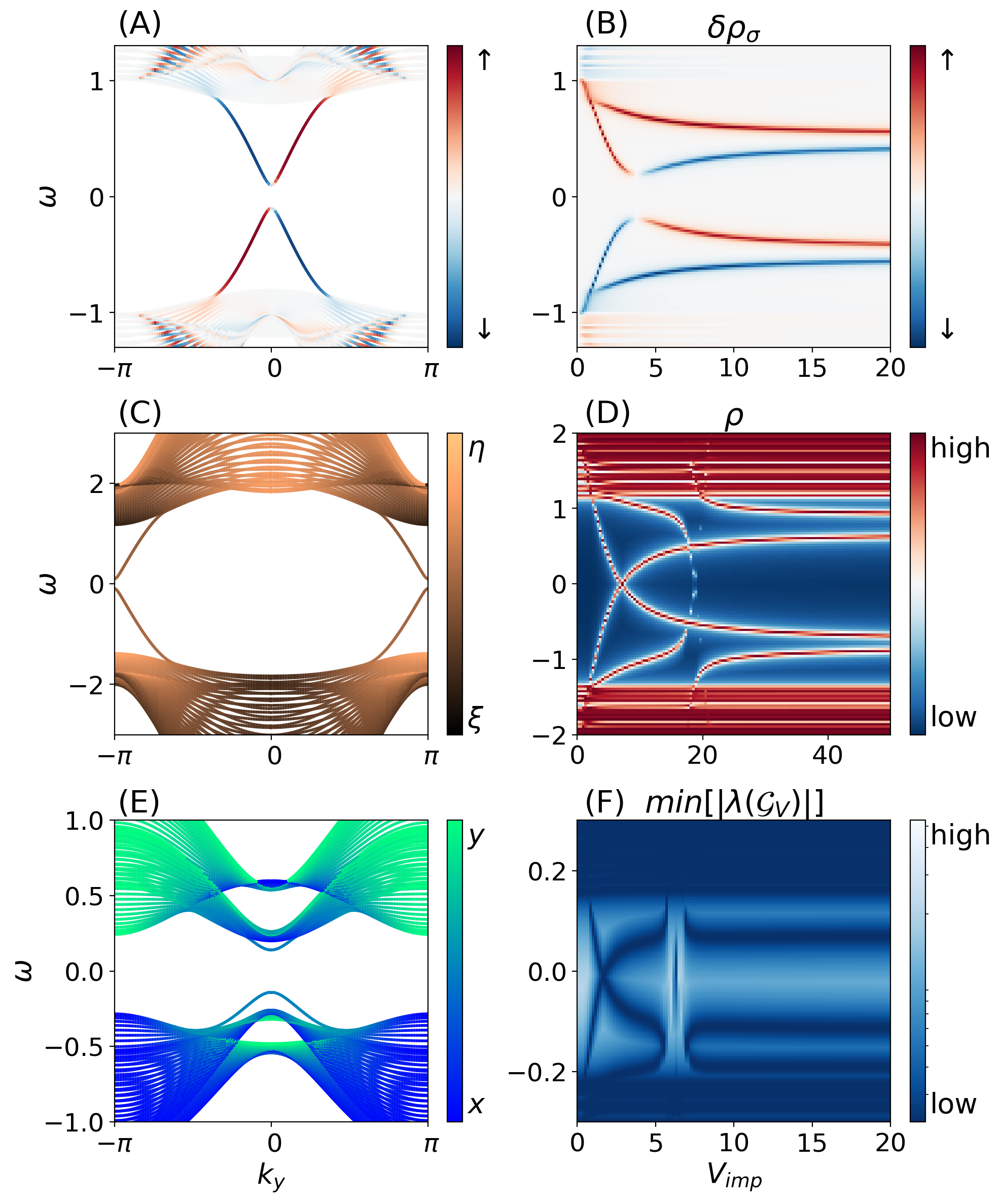}
\caption{Formation of ring states
in topologically trivial models. In the first column, we present the band structure of a ribbon with a width of 64 and a length of 128 (in the periodic direction) for three distinct models: The first line corresponds to a trivial non-interacting model with weak edge state hybridization with $\Delta_\tau=0.1t$, $M=1.0 t$ and $\alpha^{SOC}=0.05t$. In (B) we depict the corresponding spin-resolved impurity contribution to the density of states $\delta \rho_\sigma(\omega)$. The second line illustrates a trivial Mott insulator derived from the TSMBI phase by introducing a small gap in the edge states, with the resulting average density of states in the presence of an impurity $\delta \rho(\omega)$ shown in panel (D). The third line depicts a trivial Mott insulator obtained by slightly breaking time-reversal symmetry in a TSMZ phase, with $\min[|\lambda(\mathcal{G}_V)|]$ shown in panel (F). The COM parameters used to generate this figure are given in App.~\ref{App:Params}. } 
\label{fig:fig3}
\end{figure}

As discussed in Sec.\ref{sec:Ring_NI}, a Chern insulator can be topologically trivial while still exhibiting ring states with energy inside the gap, due to the importance of orbital mixing and of the bandwidth, rather than that of topology. However, this effect is not observed when solving the COM self-consistent equations~\cite{pangburn2024topological}, as the bandwidth of the interacting poles remains too large. 

To demonstrate that ring states can emerge in the self-consistent solution of a topologically trivial Mott insulator, it is easier to use other models. In particular ring states are more easily observed in models where the symmetry-protected topology is only slightly broken. The idea is that such slight symmetry breaking can be induced with minimal parameter changes, and thus ring states may persist within the energy gap. We illustrate this phenomenon using the BHZ-Hubbard model with a slight breaking of time-reversal symmetry $\mathcal{T}$.

The topological phases studied in the BHZ-Hubbard model correspond one-to-one with those of the Chern-Hubbard model. This is because the BHZ-Hubbard model is constructed by considering two time-reversal-conjugated copies of the Chern-Hubbard model. In the absence of spin-orbit coupling ($\alpha^{SOC}=0$), spin-up and spin-down sectors remain decoupled, each forming an exact replica of the Chern-Hubbard model but with opposite Chern numbers. However, even for finite spin-orbit coupling, as long as there is no gap closing of either poles or zeros, the topological properties of the Green's function can still be determined using the spin-resolved topology techniques detailed in Appendix~\ref{App:G_Spin_Topo}. Similar to the Chern-Hubbard case, three distinct topological phases can be stabilized: the Trivial Spin Mott Insulator (SMI), the Topological Spin Mott Band Insulator (TSMBI), and the Topological Spin Mott Zeros phase (TSMZ).

In Fig.~\ref{fig:fig3}, we present a summary of the impurity response results for a representative set of parameters in the BHZ-Hubbard model with finite spin-orbit coupling $\alpha^{SOC}$. The inclusion of finite spin-orbit coupling demonstrates that the observed phenomena are not fine-tuned to systems with $U(1)$ spin symmetry, highlighting the generality of the results. 
We examine three distinct phases that are trivialized by the same mechanism: slightly gapping the edge states while leaving the bulk bands largely unchanged. This is illustrated in Fig.~\ref{fig:fig3}, where in the left column we plot the bands of an infinite ribbon, calculated numerically of width $64$ and length $128$.

The first phase corresponds to the non-interacting BHZ model given by Eq.~\ref{eq:BHZ_model}, as shown in the first row of Fig.~\ref{fig:fig3}. Since the local Green's function changes only weakly when introducing a slight perturbation of the Hamiltonian, ring states persist within the energy gap. However, the model is now topologically trivial because this perturbation slightly gaps the edge states.

Next, we demonstrate that the same phenomenon occurs in Mott insulators for the bands of both poles and zeros. The second line of Fig.~\ref{fig:fig3}, is a trivial Mott insulator. It is constructed by starting from a TSMBI phase, where the poles exhibit a $\mathds{Z}_2$ topological character and the edge states acquire a small gap. Similarly to the non-interacting case, we find that this procedure does not affect the ring subgap states at very large values of the impurity potential. As discussed also for the Chern-Hubbard model, this is because this model behaves similarly to the non-interacting model for which the orbital mixing has been replaced with a holon-doublon mixing. Finally, in the third row of Fig.~\ref{fig:fig3}, a TSMZ phase is trivialized by introducing a gap in the zeros-edge states spectrum, resulting in a trivial Mott insulator for which the zeros exhibit orbital mixing. The behavior is once more similar to the non-interacting case, since the zeros can be interpreted as quasiparticles, further supporting the analogy between the behavior of the Green's function zeros and the non-interacting model.


In summary, by analyzing some topologically trivial phases of the BHZ model that exhibit significant orbital mixing and sub-gap impurity states, we confirm that, similar to the non-interacting systems, the orbital mixing serves as a reliable predictor of ring states also in strongly correlated systems.

\section{Discussion\label{sec:Discussions}}

\subsection{Summary of key findings}
Along the lines of Ref.~\cite{queiroz2024ring}, we have first reviewed the relationship between the formation of localized impurity subgap states at very large impurity potential, and the Green's function and corresponding bulk bands in non-interacting systems. 
These states are also denoted as ring states due to their specific ring-like spatial structure. The existence of these states can be closely related to both topology and the band mixing, in particular they arise most generally in topological systems, but also a finite orbital mixing, even very small, will lead to subgap ring states for a trivial system with quasi-flat bands. 

The $T$-matrix analysis used for non-interacting systems is however only applicable to models that satisfy Wick theorem. In models where the interaction term dominates, Wick's theorem is violated, and perturbation-theory-based techniques become invalid. 
In this regime, for example for a single-band Hubbard model, we have instead used a composite operator method, which gave us access to the bulk (unperturbed) Green's function. The real-space equivalent of this method gave us access also to the real-space Green's function in the presence of impurity. 

Same as for the non-interacting systems, we have tried to understand the conditions under which these states persist at very large values of the impurity potential. For this we have focused first on the implication of the composite operator method for the bulk bands of the single-band Hubbard model. We can identify these bands as corresponding to a set of weakly-interacting elementary excitations built perturbatively upon the strongly correlated ground state. The set of elementary excitations depends on the ground state, as well as the order to which the perturbation theory is extended. In this work, we focus solely on a strong local repulsion $U$, in which case at first order the set of elementary excitations consists of holons ($\xi$) and doublons ($\eta$). 
These bands can be subsequently be understood and treated in the same manner as the electronic bands are treated in the non-interacting case, and the elementary excitations $\xi$ and $\eta$ are treated as free quasiparticles with parameters that must be computed self-consistently. This approach is closely related to the methods of slave-boson mean field theory~\cite{wen1996theory}. Consequenty, the $T$-matrix formalism in the presence of an impurity applies at every iteration of the self-consistent framework. 

This suggests the emergence of a purely Mott-driven ring state where interactions alone are responsible for its formation. Such a "Mott ring state" emerges here not from orbital mixing but rather from the hybridization of holons and doublons. As long as the intra-orbital Coulomb repulsion $U$ remains finite, holons and doublons retain a finite hybridization. Furthermore, due to the Pauli exclusion principle, holons and doublons behave as orthogonal orbitals~\cite{avella2011composite}. 

Our calculations are confirming this intuition, and we have found that, same as for the non-interacting systems, the ring states can appear both in a topological system, as well as in a trivial system with band mixing. The same argument as for the non-interacting system applies for a single-band Hubbard model, thus if the lower and upper Hubbard bands exhibit a nearly flat dispersion, a ring state will emerge within the gap at large impurity strength, driven by holon-doublon hybridization. This can also be extended to multi-band trivial Mott insulators. 

Moreover, the composite operator method allowed also to observe the formation of bands of zeros in $k$-space~\cite{weinberg1965evidence} in the bulk Green's function. In this framework the Green's function zeros could be interpreted as quasiparticles~\cite{fabrizio2022emergent,blason2023unified,pasqua2024exciton,pasqua2024fermi}.
We have found that the Green's function zeros bands behave also similarly to non-interacting bands, in that they give rise to ring states (``Mott zeros ring states'') in the presence of band mixing. Thus they arise for either  bands of zeros with topological character (and automatic band mixing), as well as for non-topological bands with an almost flat dispersion and band-mixing.

\subsection{Outlook : Generalized atomic Hamiltonians}

This work, together with Ref.~\cite{pangburn2024topological}, offers a simple and unified framework for describing topological Mott insulator at the single-particle level, both in Fourier space and real space. The main limitation of this approach is its restriction to an operator basis consisting solely of holons and doublons, thereby missing effects beyond single-particle physics. A promising and computationally feasible direction to explore topological phenomena beyond the single-particle framework is to investigate multi-orbital models, incorporating non-perturbatively not only intra-orbital interactions but also inter-orbital interactions, such as those described by the Hubbard-Kanamori Hamiltonian~\cite{vojta2010orbital}.



\section{Acknowledgement}
The authors thank Q. Si and B.J. Wieder for useful discussions. A.B. and C.P. acknowledges funding from CEPIFRA (Grant No. 6704-3). The calculations are performed on the IPhT Kanta cluster. 

%


\appendix

\begin{onecolumngrid}

\vspace{3\baselineskip}

The Appendix of this paper is organized as follows. In App.\ref{App:T_matrix}, we provide additional details on the $T$-matrix formalism for non-interacting models. In App.\ref{App:Ring_Flat}, we establish the relationship between orbital mixing and ring states for a non-interacting flat-band model, using the flattening procedure outlined in App.~\ref{App:FlattenH}. Next, we focus on strongly correlated models, beginning with a review of the methods for diagnosing the topology of the two-point Green's function in App.\ref{App:G_Topo_Review}. We then provide a detailed presentation of the composite operator method, along with all the necessary equations, in App.\ref{App:Review_COM}. Finally, the composite operator parameters used throughout this work are provided in App.~\ref{App:Params}.

\section{T-matrix formalism\label{App:T_matrix}}
In this appendix, we provide a more detailed discussion of the $T$-matrix, which enables an exact solution of the single-impurity problem in non-interacting systems and was briefly introduced in Sec.~\ref{sec:T_matrix_review}.\\
In translationally invariant, non-interacting systems, the Green's function solution for a system with a localized impurity can be exactly obtained using the $T$-matrix formula~\cite{balatsky2006impurity}. The Hamiltonian corresponding to a localized impurity  at a specific lattice site $i$ is is given by
\begin{align}
    \mathcal{H}_V=  V_{\alpha\sigma} \hat{n}_{i \alpha \sigma },
\end{align}
here $V_{\alpha\sigma}$ represents the impurity strength for orbital $\alpha$ and spin $\sigma$. In Fourier space, positioning the impurity at the origin of the lattice, it reads
\begin{align}
&H_{V}=\sum\limits_{\mathbf{k},\mathbf{k}'}V_{\alpha\sigma}c^\dagger_{\mathbf{k}\alpha\sigma}c_{\mathbf{k}'\alpha\sigma}&,
\label{eq:H_Impurity}
\end{align}
It can then be demonstrated through the resummation of Feynman diagrams that the exact Green's function of the problem $\mathcal{G}_{imp}$ is given by
\begin{align}
&\mathcal{G}_{imp}(\mathbf{k},\mathbf{k}',\omega)=\delta_{\mathbf{k},\mathbf{k}'}\mathcal{G}_0(\mathbf{k},\omega)+\mathcal{G}_0(\mathbf{k},\omega)T(\omega)\mathcal{G}_0(\mathbf{k}',\omega) &\\[5pt]
&T(\omega)=\Big(\mathds{1}-\hat{V}\sum\limits_{\mathbf{k}} \mathcal{G}_0(\mathbf{k},\omega)\Big)^{-1}\hat{V}&
\label{eq:TMatrixFormula}
\end{align}
The resulting formula utilizes Wick's theorem, which states that the correlation function of an even number of fermionic operators can be decomposed into a sum of products of bilinear correlation functions. $T(\omega)$ is called the $T$-matrix and encodes the effects of the impurity.

The important insight given by Eq.~\ref{eq:TMatrixFormula} is that impurity states originate from the poles of the $T$-matrix, making it the central quantity to analyze. The mathematical condition to have a bound state at an energy $\omega$ for an impurity potential $\hat{V}$ then read

\begin{align}
&\det\Big[\mathds{1}-\hat{V}\sum\limits_{\mathbf{k}}\mathcal{G}_0(\mathbf{k},\omega)\Big]=0&
\label{eq:GreenFunctionSubgap}
\end{align}
In the limit where the strength of the impurity is very large, $||\hat{V}||\rightarrow \infty$, this can be rephrased as 
\begin{align}
&\det\Big[\mathcal{G}_0(\mathbf{R}=\mathbf{0},\omega)\Big]=0&
\label{eq:GreenFunctionZeros}
\end{align}
The key observation is that this condition is independent of the impurity potential, implying that it solely probes the properties of the clean system.  

\section{Subgap states for large impurity non-interacting case\label{App:Ring_Flat}}

In this appendix, we explore why subgap impurity states can survive in non-interacting systems for considerable impurity strength. For such systems the T-matrix formalism is exact and the following condition must hold:
\begin{align}
    \det \left[ \sum_k \mathcal{G}(k,\omega) \right] = 0.
\end{align}
We refer to the sum of the Green's function over the whole Brillioun zone, i.e. $\mathcal{G}(\bf{R=0,\omega})=\sum_k \mathcal{G}(k,\omega)$ the local Green's function. For a general two band model the Green's function is a $2\times 2$ matrix and the components are given  by $\mathcal{G}_{\alpha \beta}$ where $\alpha$ and $\beta$ are the orbital indices.

In gapped systems the density of states, $\rho(\omega)=-\frac{1}{\pi} \sum_{k,\alpha} \text{Im } \mathcal{G}_{\alpha \alpha}(\mathbf{k},\omega)$ vanishes for $ \vert \omega \vert <\Delta$, where $\Delta$ is the magnitude of the gap. The Green's function can be written using spectral decomposition with eigenvectors $\mathbf{\Lambda}_n (\mathbf{k})$ and corresponding eigenvalues $\epsilon_n(\mathbf{k})$ as
\begin{align}
\mathcal{G}_{\alpha\beta}(\mathbf{k},\omega)=\sum\limits_{n}\dfrac{\Lambda^\dagger_{\alpha n}(\mathbf{k})\Lambda_{n\beta}(\mathbf{k})}{\omega-\epsilon_n(\mathbf{k})+i\delta}
\end{align}
The diagonal Green's function becomes
\begin{align}
    \mathcal{G}^I_{\alpha\alpha}(\mathbf{k},\omega)&=-\pi\sum\limits_n \vert \Lambda_{\alpha n} (\mathbf{k}) \vert^2 \delta(\omega-\epsilon_n(\mathbf{k}))\\
    \mathcal{G}^R_{\alpha\alpha}(\mathbf{k},\omega)&=\sum\limits_n \frac{\vert \Lambda_{\alpha n} (\mathbf{k}) \vert^2}{\omega - \epsilon_n(k)}
\end{align}
where $\mathcal{G}^I$ and $\mathcal{G}^R$ are the imaginary and real part of the diagonal Green's function. For energies inside the gap $\mathcal{G}_{\alpha \alpha}^I=0$. Next we focus on the off-diagonal part of the Green's function $\mathcal{G}_{\alpha\beta}$ inside the gap, and we get
\begin{align}
\mathcal{G}^I_{\alpha\beta}(\mathbf{k},\omega)&=\sum\limits_n\text{Im}\left[\Lambda^\dagger_{\alpha n}(\mathbf{k})\Lambda_{n\beta}(\mathbf{k})\right]\mathcal{P}\left(\dfrac{1}{\omega-\epsilon_n(\mathbf{k})}\right)\\
\mathcal{G}^R_{\alpha\beta}(\mathbf{k},\omega)&=\sum\limits_n\text{Re}\left[\Lambda^\dagger_{\alpha n}(\mathbf{k})\Lambda_{n\beta}(\mathbf{k})\right]\mathcal{P}\left(\dfrac{1}{\omega-\epsilon_n(\mathbf{k})}\right)
\end{align}
We apply the condition of finding zero of the local Green's function by using Eq.~\ref{eq:GreenFunctionZeros},
\begin{align}
    \sum_{\mathbf{k}} \mathcal{G}^R_{ss}(\mathbf{k},\omega)  \sum_{\mathbf{k}} \mathcal{G}^R_{pp}   (\mathbf{k},\omega) -\sum_{\mathbf{k}}  \mathcal{G}_{sp}(\mathbf{k},\omega) \sum_{\mathbf{k}}  \mathcal{G}_{ps}(\mathbf{k},\omega)=0
    \label{eq:AppBeq1}
\end{align}
We subsequently make an assumption that the interorbital Green's function over the entire Brillouin zone vanishes i.e. $\sum_\mathbf{k} \mathcal{G}_{\alpha\beta}=0$. This naturally follows for the interorbital hoppings of the Chern insulator model and the BHZ model. Hence, the condition for local Green's function zero reduces to 
\begin{align}
        \sum_{\mathbf{k}} \mathcal{G}^R_{\alpha \alpha}(\mathbf{k},\omega)  &=0.
\end{align}
Next, we project the dispersion onto a flat band without changing the eigenvectors, which allows us to conduct analytical calculations. The detailed procedure for this is presented in App.~\ref{App:FlattenH}. The flat-band diagonal Green's function becomes $\tilde{\mathcal{G}}_{\alpha\alpha}$, expressed in $\mathbf{k}$-space as  
\begin{align}
\label{eq:AppBeq2}
\tilde{\mathcal{G}}^R_{\alpha\alpha}(\mathbf{k},\omega) = \sum\limits_n \dfrac{\vert \Lambda_{\alpha n} (\mathbf{k}) \vert^2}{\omega - \epsilon_n}
\end{align}
where $\epsilon_n$ is independent of \(\mathbf{k}\). Replacing Eq.~\ref{eq:AppBeq2} into Eq.~\ref{eq:AppBeq1} , we derive the conditions for a local Green's function zero within the gap:  
\begin{align}
\frac{1 - \chi_\alpha}{\omega - \epsilon_1} + \frac{\chi_\alpha}{\omega - \epsilon_2} = 0
\end{align}
where the gap is defined as \(\Delta = \epsilon_2 - \epsilon_1\). We also substitute \(\sum_{\mathbf{k}} \vert \Lambda_{1\alpha} (\mathbf{k}) \vert^2 = 1 - \chi_\alpha\) and \(\vert \Lambda_{2\alpha} (\mathbf{k}) \vert^2 = \chi_\alpha\), with \(0 \leq \chi_\alpha \leq 1\). In other words, the factors in the numerators represent the overlap of the \(\alpha\)-orbital with the \(n\)-th band. If the bands are purely of one orbital character such that \(\chi_\alpha = 0\), then local Green's function zeros cannot exist within the gap. This corresponds to the case of trivial band insulators with no mixed orbitals. However, if there is a weak mixing of orbital characters, the existence of a zero is described by  
\begin{align}
\omega^{(0)}_\alpha = \epsilon_2 - \Delta \chi_\alpha .
\label{Eq:zero_ana}
\end{align}
Therefore, the bands do not need to be topological to exhibit stable subgap states within the gap. If the bands possess mixed orbital character and the interorbital Green's function vanishes over the entire Brillouin zone, this condition is sufficient for stable subgap states at large impurity strength. In general, the subgap states at large impurity strength probe the mixed orbital character of the bands with interorbital hybridization rather than serving as a direct probe of topological bands. In Fig.~(\ref{fig:Appfig1}) we numerically test our prediction for the position of the subgap state at large $V_{\rm imp} $ for a flat-band spectrum, which matches our analytical calculations from Eq.~(\ref{Eq:zero_ana}).

\begin{figure}[h!]
\includegraphics[width=10cm]{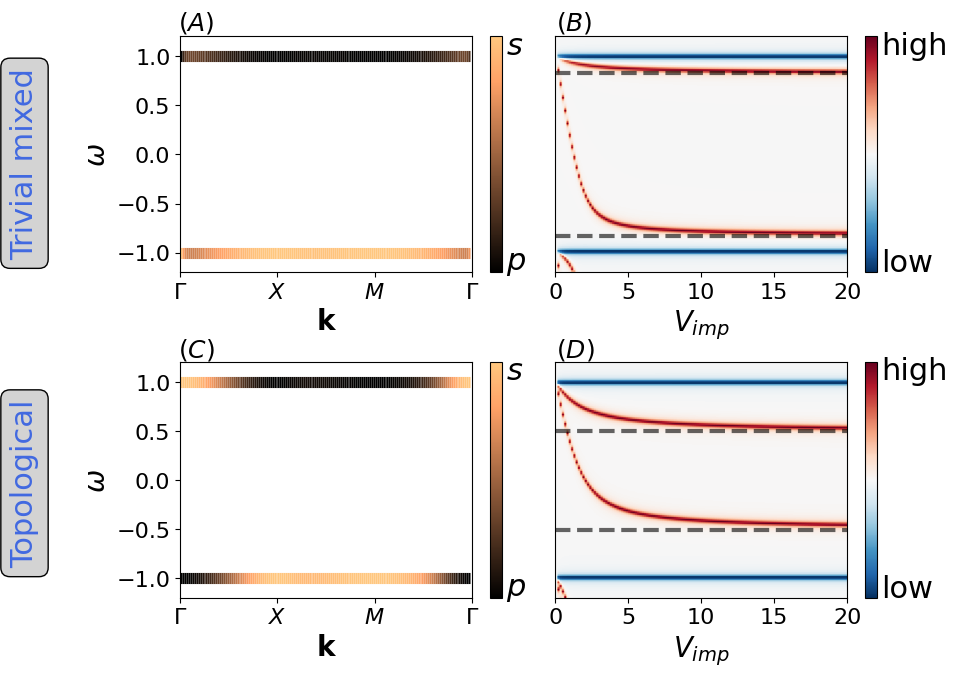}
\caption{Comparison of numerical and analytical predictions for the positions of impurity-induced subgap states in flat-band systems. In panels (A) and (C), orbitally-resolved bands are displayed for (A) topological bands and (C) trivial bands with mixed orbital character. This corresponds to a flattened two-band Chern insulator with $M=1.0t$ (A) and $M=2.05t$ (C). In panels (B) and (D), the correction to the density of states is shown as a function of impurity strength for the corresponding bands. The analytical prediction for the asymptotic position of the subgap state at large impurity strength, indicated by the dashed line, aligns precisely with the numerical computation.} 
\label{fig:Appfig1}
\end{figure}

\section{Flattened Hamiltonian\label{App:FlattenH}}
In this appendix, we briefly review the procedure for flattening any non-interacting Hamiltonian, which involves preserving its eigenvectors while rendering the eigenspectrum dispersionless.\\
Let's consider a non-interacting Hamiltonian $\mathcal{H}$,
\begin{align}
&\mathcal{H}=\sum\limits_{\mathbf{k}\alpha\beta}c^\dagger_{\mathbf{k}\alpha} \left[H_0(\mathbf{k})\right]_{\alpha\beta}c_{\mathbf{k}\beta},&
\end{align}
where $c^\dagger_{\mathbf{k}\alpha}$ and $c_{\mathbf{k}\alpha}$ are creation and annihilation operators and $H_0(\mathbf{k})$ is the Bloch Hamiltonian. Let's denote $|u_n(\mathbf{k})\rangle$ the eigenvectors of $H_0(\mathbf{k})$. We consider a model with a gap between the occupied and unoccupied bands, denoted as $occ$ and $unocc$, respectively. Then we can construct the projector $P(\mathbf{k})$ onto the occupied bands at each momentum as
\begin{align}
&P(\mathbf{k})=\sum\limits_{n\in occ} |u_n(\mathbf{k})\rangle \langle u_n(\mathbf{k})|&
\end{align}
Using $P(\mathbf{k})$ we can construct the following operators
\begin{align}
&Q(\mathbf{k})=\mathds{1}-2P(\mathbf{k})&
\end{align}
We notice that both $Q(\mathbf{k})$ and $H_0(\mathbf{k})$ have the same eigenvectors $|u_n(\mathbf{k})\rangle$, however $Q(\mathbf{k})$ has only $\pm 1$ eigenvalues. If the occupied bands of $Q(\mathbf{k})$ are topological, then $Q(\mathbf{k})$ cannot be constructed using strictly local hopping~\cite{chen2014impossibility}, implying that it must inherently correspond to an infinite-range tight-binding Hamiltonian. $Q(\mathbf{k})$ is useful because it preserves the same topology as $H_0(\mathbf{k})$ while allowing for  analytical computations.

\section{Single-particle Green's function topology \label{App:G_Topo_Review}}
In this appendix, we review the methods for studying the topology of a system bases on the single-particle Green's function.
\subsection{Winding number $N_3[\mathcal{G}]$}

In non-interacting models, the strong topological invariants described by the tenfold way~\cite{chiu2016classification} can be calculated straightforwardly. For strongly interacting models, the ground state is generally unknown, making the diagnosis of topology more challenging. Topological invariants can still be determined by introducing a periodic flux and counting the number of states that transition during its insertion~\cite{oshikawa2000topological}. Another recently popular approach to describe topology~\cite{misawa2022zeros,wagner2023mott,bollmann2024topological,chen2024dirac,setty2024electronic,setty2024symmetry} involves using Green's functions, which can be computed through various numerical methods. Among these, the single-particle Green's function is the most straightforward. For a $2d$ model,  the topology of a single-particle Green's function $\mathcal{G}(\mathbf{k},\omega)$ can be described using the following winding number formula:
\begin{align}
N_3\big[\mathcal{G}\big]=&\sum_{\alpha \beta \gamma}\dfrac{\epsilon_{\alpha\beta\gamma}}{6}\int_{-\infty}^{+\infty}d\omega \int \dfrac{d^2k}{(2\pi)^2} \text{Tr}\big[(\mathcal{G}^{-1}\partial_{k_\alpha}\mathcal{G})(\mathcal{G}^{-1}\partial_{k_\beta}\mathcal{G})(\mathcal{G}^{-1}\partial_{k_\gamma}\mathcal{G})\big].&
\label{eq:N3_winding}
\end{align}

For $2d$ non-interacting models, $N_3[\mathcal{G}]$ directly corresponds to the Chern number. However, this relationship breaks down in correlated systems~\cite{peralta2023connecting}. When the ground state is degenerate, as in topologically ordered systems like the fractional quantum Hall effect, the correspondence no longer holds~\cite{gurarie2013topological}. Additionally, the formula differs from the Chern number in models where the electronic spectral weight is split, such as in Mott insulators~\cite{zhao2023failure}, which display bands of Green's function zeros. A phase is considered topological at the single-particle level when $N_3[\mathcal{G}]$ is non-zero. This topology can originate from either the poles (regular bands) or from the $k$-space zeros of the Green's function. Although zeros do not contribute to the electromagnetic response, and their associated topological states may not be directly observable, zero-energy bands remain mathematically classifiable.

\subsection{Spin-resolved winding number $N^S_3[\mathcal{G}]$\label{App:G_Spin_Topo}}
In the class $AII$ model, protected by spinful time-reversal symmetry, the Chern number is trivial, which also results in the triviality of the winding number $N_3[\mathcal{G}]$. Similar to non-interacting insulators, non-trivial spin-resolved topological invariants can be defined to characterize the presence of protected edge states. \\

In the case where there is $U(1)$-$S_z$ spin conservation, the Green function can be expressed in a block-diagonal form in the spin basis
\begin{align}
&\mathcal{G}^{S_z}(\mathbf{k},\omega)=\begin{pmatrix}
\mathcal{G}_{\uparrow\uparrow}(\mathbf{k},\omega) & 0 \\
0 & \mathcal{G}_{\downarrow\downarrow}(\mathbf{k},\omega)\\
\end{pmatrix}&
\end{align}

We can then compute the winding number of both $\mathcal{G}_{\uparrow\uparrow}$ and $\mathcal{G}_{\downarrow\downarrow}$ through
\begin{align}
&N_3\big[\mathcal{G}_{\sigma\sigma}\big]=\sum_{\alpha \beta \gamma}\dfrac{\epsilon_{\alpha\beta\gamma}}{6}\int_{-\infty}^{+\infty}d\omega \int \dfrac{d^2k}{(2\pi)^2}  \text{Tr}\big[(\mathcal{G}_{\sigma\sigma}^{-1}\partial_{k_\alpha}\mathcal{G}_{\sigma\sigma})(\mathcal{G}_{\sigma\sigma}^{-1}\partial_{k_\beta}\mathcal{G}_{\sigma\sigma})(\mathcal{G}_{\sigma\sigma}^{-1}\partial_{k_\gamma}\mathcal{G}_{\sigma\sigma})\big]&
\label{eq:N3_winding}
\end{align}
In phases preserving time-reversal symmetry, where $N_3\left[\mathcal{G}^{S_z}\right]$ vanishes, a non-trivial $\mathds{Z}_2$ topological invariant, $\mathcal{C}_s$, can still be defined as
\begin{align}
&\mathcal{C}_s=\left(N_3[\mathcal{G}_{\uparrow\uparrow}]-N_3[\mathcal{G}_{\downarrow\downarrow}]\right)/2 \quad \mod 2&
\end{align}

This previous definition relies on $U(1)$ spin symmetry. In non-interacting models with spin-orbit coupling, or in interacting materials with non-paramagnetic magnetic correlations, the $U(1)$ spin symmetry is broken, rendering the above procedure invalid. In the non-interacting case, if the spin operator $\hat{n}\cdot\hat{S}$ is approximately conserved, the $P(\mathbf{k})\hat{n}\cdot\hat{S}P(\mathbf{k})$ spectrum can be computed with $P(\mathbf{k})$ the projector onto the occupied bands. This enables the extension of the computation of $\mathds{Z}_2$ invariant to models where the nonzero eigenvalues of $P(\mathbf{k})\hat{n}\cdot\hat{S}P(\mathbf{k})$  are gapped, allowing for a clear distinction between spin-up-like and spin-down-like states. Moreover, this formalism can be further generalized to characterize the topology of spinful single-particle Green's functions. \\

In the presence of spin-orbit coupling, $U(1)$ spin-symmetry is broken and the single-particle Green's function can in general be expressed as
\begin{align}
&\mathcal{G}(\mathbf{k},\omega)=\begin{pmatrix}
\mathcal{G}_{\uparrow\uparrow}(\mathbf{k},\omega) & \mathcal{G}_{\uparrow\downarrow}(\mathbf{k},\omega) \\
\mathcal{G}_{\downarrow\uparrow}(\mathbf{k},\omega) & \mathcal{G}_{\downarrow\downarrow}(\mathbf{k},\omega) \\
\end{pmatrix}&
\end{align}

Because $\mathcal{G}$ is no longer block-diagonal, the direct computation of $N_3[\mathcal{G}_{\sigma\sigma}]$ is no longer possible. Following~\cite{setty2023topological}, we can study some frequency dependent topological invariant. To do so, we fix the frequency $\omega$. One can then diagonalize $\mathcal{G}(\mathbf{k},\omega)$
\begin{align}
&\mathcal{G}(\mathbf{k},\omega)|v_i(\mathbf{k},\omega)\rangle = \lambda_i(\mathbf{k},\omega)|v_i(\mathbf{k},\omega)\rangle,&
\end{align}
where $i\in \{1,\cdots,N_{orb} \}$ indexes the eigenvectors of $\mathcal{G}$.
The projector $P_\mathcal{G}(\mathbf{k},\omega)$ onto the occupied bands of $\mathcal{G}(\mathbf{k},\omega)$ can then be written as:
\begin{align}
&P_{\mathcal{G}}(\mathbf{k},\omega)=\sum\limits_{i\in occ}|v_i(\mathbf{k},\omega)\rangle\langle v_i(\mathbf{k},\omega)|.&
\end{align}
Using $P_\mathcal{G}$, one can construct the $P_\mathcal{G}\hat{n}\cdot\hat{S}P_\mathcal{G}$ spectrum by diagonalizing this operator. After removing the zero eigenvalues of $\mathcal{P}_\mathcal{G}$, if the spectrum $P_\mathcal{G}\hat{n}\cdot\hat{S}P_\mathcal{G}$ is gapped throughout all the Brillouin zone, then one can construct the projector on the lower and upper spin bands. In the limit of an exact $U(1)$ spin symmetry, the eigenvalues of $\mathcal{P}\hat{n}\cdot\hat{S}$ are restricted to $+1$ or $-1$. However, when $U(1)$ spin symmetry is broken, these eigenvalues vary continuously within the range $[-1,+1]$. 
If there is a spin gap, one can define the projectors $P_-(\mathbf{k},\omega)$ and $P_+(\mathbf{k},\omega)$ on the lower and upper spin bands respectively. One can then compute the topological invariant associated with $P_-$ and $P_+$. In $2d$, one can compute the frequency dependent Chern number $N_\pm(\omega)$.

In the limit of an exact $U(1)$ spin symmetry, the eigenvalues of $\mathcal{P}\hat{n}\cdot\hat{S}$ are restricted to $\pm 1$. However, when $U(1)$ spin symmetry is broken, these eigenvalues continuously vary within the range $[-1,+1]$. If a spin gap is present, one can define the projectors  $P_-(\mathbf{k},\omega)$ and $P_+(\mathbf{k},\omega)$  onto the lower and upper spin bands, respectively. This allows for the computation of the topological invariants associated with $P_-$ and $P_+$ . In two dimensions, the frequency-dependent Chern number $N_\pm(\omega)$ can then be evaluated.

\begin{align}
&N_\pm(\omega)=\dfrac{1}{2\pi i}\int_0^{2\pi}\int_0^{2\pi}d^2\mathbf{k}\text{Tr}\Big(P_\pm(\mathbf{k},\omega)\left[\partial_{k_x}P_{\pm}(\mathbf{k},\omega),\partial_{k_y}P_\pm(\mathbf{k},\omega)\right]\Big)&
\end{align}

We call the integrand of $N_\pm(\omega)$ the Green spin-resolved Berry curvature defined as
\begin{align}
&\mathcal{B}_{\pm}(\mathbf{k},\omega)=\Big(P_\pm(\mathbf{k},\omega)\left[\partial_{k_x}P_{\pm}(\mathbf{k},\omega),\partial_{k_y}P_\pm(\mathbf{k},\omega)\right]\Big)&\\
&\mathcal{B}_{S_z}(\mathbf{k},\omega)=\mathcal{B}_+(\mathbf{k},\omega)-\mathcal{B}_-(\mathbf{k},\omega)&
\end{align}

At each frequency $\omega$ one can then define a spin-resolved topological number $\mathcal{C}^{\mathcal{G}}_s(\omega)$ as 
\begin{align}
&\mathcal{C}^{\mathcal{G}}_s(\omega)=\left[N_+(\omega)-N_-(\omega)\right]/2 \ \text{mod} 2&
\end{align}
\section{Details of the composite-operator method\label{App:Review_COM}}
The composite-operator method (COM)~\cite{avella2011composite} has been recently reviewed~\cite{Haurie_2024,banerjee2024charge,pangburn2024spontaneous}. We provide here a quick review of the method and all the important equations necessary to reproduce our results. The COM approach involves identifying a set of operators that are the long-lived excitations above of a given ground state. In the following we consider the holon $\xi_{i,\tau\sigma}=c_{i\tau\sigma}\left(1-n_{i,\tau\overline{\sigma}}\right)$ and doublon $\eta_{i,\tau\sigma}=c_{i\tau\sigma}n_{i,\tau\overline{\sigma}}$ as good excitations for a Mott insulator with $\sigma$ the spin and $\tau$  the orbital degree of freedom. Here, our focus is not on verifying the validity of this statement but rather on exploring the physical consequences of this assumption. \\

The COM method is based on the equation of motion of these operators, termed "composite" because they do not follow the standard fermionic anti-commutation relations. Note that in this formalism the electrons splits into a holon and a doublon excitations such that $c=\xi+\eta$ and hence single particle electronic Green's function can be constructed from the composite Green's functions. Given an arbitrary composite operator basis $\mathbf{\Psi}$, the composite Green's function is defined as
\begin{align}
    \mathds{G}_{ij}(t)=-i\Theta(t) \left\langle \{ \mathbf{\Psi}(t ) , \mathbf{\Psi}^\dagger(0) \} \right\rangle,
\end{align}
where $\Theta(t)$ is the Heaviside theta function. One proceeds by studying the equation of motion of $\mathds{G}(t)$. As is typical with such methods, the equations do not naturally close and some approximations have to be made. The COM method assumes that the composite current $\mathbf{J}$ remains proportional to the composite-operator basis such that ${\mathbf{J}(t)=i\left[ \mathcal{H} , \mathbf{\Psi}  \right]  (t) \approx \mathds{E} \Psi(t)}$ where $\mathds{E}$ is called the $E$-matrix and need to be computed self-consistently. Without any symmetry assumptions, the $E$-matrix is an $n_{COM}N\times n_{COM}N$ matrix, where $n_{COM}$ is the size of the COM basis at each lattice site, and $N$ the total number of lattice sites. This approximation leads to the following form for the composite operator Green's function:

\begin{align}
    \mathds{G}^{R/A}(\omega)=\left[ (\omega \pm i0^+) \mathds{1} - \mathds{E} \right]^{-1} \mathds{I},
    \label{Eq:Grealomega}
\end{align}
where $\mathds{1}$ is the $n_{COM}N\times n_{COM}N$ identity matrix. $\mathds{I}$ in the Green's 
 function equation is called the $I$-matrix is is given by
\begin{align}
 \mathds{I} = \left\langle \{ \mathbf{\Psi}(0),\mathbf{\Psi}(0)^\dagger \}  \right\rangle.\label{Eq:IMat}
\end{align}
The $I$-matrix is measuring the orthogonality of the composite operators.
Another useful matrix that can be defined is the $M$-matrix given by
\begin{align}
\mathds{M}= \left\langle \{ \mathbf{j}(0),\mathbf{\Psi}(0)^\dagger \}  \right\rangle.
\label{Eq:MMat}
\end{align}
The $M$-matrix quantifies how a given operator mixes with other operators as a function of time, providing insight into the time evolution and the interactions between different operators. The usefulness of the $M$ and $I$ matrices lies in the fact that the $E$-matrix, which represents physical quantities with eigenvalues corresponding to the energy of quasiparticle excitations, can be expressed in terms of these two matrices as follows:
\begin{align}
\mathds{E}=\mathds{M}\mathds{I}^{-1}
\label{Eq:EMat}
\end{align}

The study of local probes requires a real-space formulation of the composite operator formalism. In the following, we review for completeness the form of the $I$- and $M$-matrix for the Chern-Hubbard model and BHZ-Hubbard model.

\vspace{1\baselineskip}

The $M$-matrix and $I$-matrix depend on a set of unknown parameters, whose number is determined by the Hamiltonian and the chosen composite operator basis. A subset of these parameters, denoted as $e$, can be determined from single-particle on-site and nearest-neighbor inter-site correlation functions. The correlation matrix for the composite operator basis is obtained using the fluctuation-dissipation theorem
\begin{align}
    \mathds{C}_{\alpha \gamma}=\langle\mathbf{\Psi}_\alpha\mathbf{\Psi}^\dagger_\gamma\rangle= -\int \frac{d\omega}{4 i \pi} \left[1+\tanh\left(\frac{\beta \omega}{2}\right) \right] \left(\mathds{G}^R(\omega))-\mathds{G}^A(\omega)\right)_{\alpha \gamma}.
    \label{Eq:FullCorr}
\end{align}
with $\beta$ the inverse temperature and $\alpha \gamma$ general composite Green's function indices. \\
Other parameters, denoted as $p$, cannot be directly computed from the composite correlation matrix $\mathds{C}_{\alpha\gamma}$ as they involve higher-order correlation functions. To determine them self-consistently, an appropriate decoupling scheme must be employed, which is reviewed in App.~\ref{App:Roth}.

\subsection{Chern-Hubbard model}

In the Chern-Hubbard model, the orbital degree of freedom $\tau$ is introduced, with $\tau\in\{s,p\}$. We first choose the following $4N$-component basis for the composite operator, where $N$ indicates the number of lattice sites. As we are using the paramagnetic assumption, $\sigma$ here can be either $\uparrow$ or $\downarrow$.
\begin{align}
\label{eq:BasisHubbard}&\mathbf{\Psi}=\left(\xi_{1s\sigma},\cdots, \xi_{Ns\sigma}, \eta_{1s\sigma},\cdots, \eta_{Ns\sigma}, \xi_{1p\sigma},\cdots, \xi_{Np\sigma}, \eta_{1p\sigma},\cdots, \eta_{Np\sigma}\right)^T&
\end{align}

\subsubsection{I-matrix}

The components of the $I$-matrix in this basis are given by
\begin{align}
& \mathds{I}_{i,j}=\delta_{i,j}\left(1-n_s(i)/2\right) &\\
& \mathds{I}_{i+N,j+N}=\delta_{i,j}n_s(i)/2 &\\
& \mathds{I}_{i+2N,j+2N}=\delta_{i,j}\left(1-n_p(i)/2\right) &\\
& \mathds{I}_{i+3N,j+3N}=\delta_{i,j}n_p(i)/2 &
\end{align}
where $n_s(i)$ and $n_p(i)$ are the average electronic density a site $i$.

\subsubsection{M-matrix}
The intra-orbital $M$-matrix elements for the Chern model are given by
\begin{align}
    &\mathds{M}_{i,j} = -\delta_{ij} \left[(\mu+M) \left(1-\frac{n_s(i)}{2}\right) + te_{s}(i) - te_{sp}(i)\right] 
    - t \left(1-\frac{n_s(i) + n_s(j)}{2} + p_{ss}(i,j)\right) \label{eq:M11} \\
    &\mathds{M}_{i,j+N} = \delta_{ij}\left[te_s(i) - te_{sp}(i)\right] 
    - t \left(\frac{n_s(j)}{2} - p_{ss}(i,j)\right) \label{eq:M12} \\
    &\mathds{M}_{i+N,j} = \delta_{ij}\left[te_s(i) - te_{sp}(i)\right] 
    - t \left(\frac{n_s(i)}{2} - p_{ss}(i,j)\right) \label{eq:M21} \\
    &\mathds{M}_{i+N,j+N} = -\delta_{ij} \left[\left(\mu+M - U_s\right) \frac{n_s(i)}{2} + te_s(i) - te_{sp}(i)\right] 
    - t p_{ss}(i,j) \label{eq:M22} \\
    &\mathds{M}_{i+2N,j+2N} = -\delta_{ij} \left[\mu-M \left(1-\frac{n_p(i)}{2}\right) - te_p(i) + te_{ps}(i)\right] 
    + t \left(1 - \frac{n_p(i) + n_p(j)}{2} + p_{pp}(i,j)\right) \label{eq:M33} \\
    &\mathds{M}_{i+2N,j+3N} = \delta_{ij} \left[-te_p(i) + te_{ps}(i)\right] 
    + t \left(\frac{n_p(j)}{2} - p_{pp}(i,j)\right) \label{eq:M34} \\
    &\mathds{M}_{i+3N,j+2N} = \delta_{ij} \left[-te_p(i) + te_{ps}(i)\right] 
    + t \left(\frac{n_p(i)}{2} - p_{pp}(i,j)\right) \label{eq:M43} \\
    &\mathds{M}_{i+3N,j+3N} = -\delta_{ij} \left[\left(\mu-M - U_p\right) \frac{n_p(i)}{2} - te_p(i) + te_{ps}(i)\right] 
    + t p_{pp}(i,j) \label{eq:M44}
\end{align}

We define $\delta_x$ and $\delta_y$ as vectors connecting one site to its neighboring sites in the $\hat{x}$ and $\hat{y}$ directions, respectively. The inter-orbital matrix elements for $j=i+ \delta_x$ can be written as:

\begin{align}
    &\mathds{M}_{i,i+\delta_x+2N} = a(i,i+\delta_x) t\left(1-\dfrac{n_s(i)+n_p(i+\delta_x)}{2}+p_{sp}(i,i+\delta_x)\right) \label{eq:M13x} \\
    &\mathds{M}_{i+2N,i+\delta_x} = a(i,i+\delta_x) t\left(1-\dfrac{n_s(i+\delta_x)+n_p(i)}{2}+p_{sp}(i,i+\delta_x)\right) \label{eq:M31x} \\
    &\mathds{M}_{i,i+\delta_x+3N} = a(i,i+\delta_x) t\left(\dfrac{n_p(i+\delta_x)}{2}-p_{sp}(i,i+\delta_x)\right) \label{eq:M14x} \\
    &\mathds{M}_{i+3N,i+\delta_x+N} = a(i,i+\delta_x) t\left(\dfrac{n_p(i)}{2}-p_{sp}(i,i+\delta_x)\right) \label{eq:M41x} \\
    &\mathds{M}_{i+N,i+\delta_x+2N} = a(i,i+\delta_x) t\left(\dfrac{n_s(i)}{2}-p_{sp}(i,i+\delta_x)\right) \label{eq:M23x} \\
    &\mathds{M}_{i+2N,i+\delta_x+N} = a(i,i+\delta_x) t\left(\dfrac{n_s(i+\delta_x)}{2}-p_{sp}(i,i+\delta_x)\right) \label{eq:M32x} \\
    &\mathds{M}_{i+N,i+\delta_x+3N} = a(i,i+\delta_x) tp_{sp}(i,i+\delta_x) \label{eq:M24x} \\
    &\mathds{M}_{i+3N,i+\delta_x+N} = a(i,i+\delta_x) tp_{sp}(i,i+\delta_x) \label{eq:M42x}
\end{align}
with $a(i,i+\hat{x})=i$ and $a(i,i-\hat{x})=-i$. For $j=i+\delta_y$, the inter-orbital matrix elements are given by

\begin{align}
    &\mathds{M}_{i,i+\delta_y+2N} = -a(i,i+\delta_y) t\left(1-\dfrac{n_s(i)+n_p(i+\delta_y)}{2}+p_{sp}(i,i+\delta_y)\right) \label{eq:M13y} \\
    &\mathds{M}_{i+2N,i+\delta_y} = a(i,i+\delta_y) t\left(1-\dfrac{n_s(i+\delta_y)+n_p(i)}{2}+p_{sp}(i,i+\delta_y)\right) \label{eq:M31y} \\
    &\mathds{M}_{i,i+\delta_y+3N} = -a(i,i+\delta_y) t\left(\dfrac{n_p(i+\delta_y)}{2}-p_{sp}(i,i+\delta_y)\right) \label{eq:M14y} \\
    &\mathds{M}_{i+3N,i+\delta_y+N} = a(i,i+\delta_y) t\left(\dfrac{n_p(i)}{2}-p_{sp}(i,i+\delta_y)\right) \label{eq:M41y} \\
    &\mathds{M}_{i+N,i+\delta_y+2N} = -a(i,i+\delta_y) t\left(\dfrac{n_s(i)}{2}-p_{sp}(i,i+\delta_y)\right) \label{eq:M23y} \\
    &\mathds{M}_{i+2N,i+\delta_y+N} = a(i,i+\delta_y) t\left(\dfrac{n_s(i+\delta_y)}{2}-p_{sp}(i,i+\delta_y)\right) \label{eq:M32y} \\
    &\mathds{M}_{i+N,i+\delta_y+3N} = -a(i,i+\delta_y) tp_{sp}(i,i+\delta_y) \label{eq:M24y} \\
    &\mathds{M}_{i+3N,i+\delta_y+N} = a(i,i+\delta_y) tp_{sp}(i,i+\delta_y) \label{eq:M42y}
\end{align}
with $a(i,i+\hat{y})=1$ and $a(i,i-\hat{y})=-1$.

\subsubsection{Self-consistent 
parameters}
Since the self-consistent parameters of the Chern-Hubbard model overlap with those of the BHZ-Hubbard model,  we provide the definitions of the $e$ and $p$ parameters in terms of correlation functions in App.~\ref{App:BHZ_SC_params}.

\subsection{BHZ-Hubbard model}

In the BHZ-Hubbard model, both the orbital degrees of freedom $\tau \in \{s,p\}$ and the spin degrees of freedom $\sigma\in\{\uparrow,\downarrow\}$ explicitly appear in the equations as we relax the paramagnetic assumption. The composite operator basis now expands to a size of $8N$, where $N$ is the number of lattice sites.
\begin{align}
\mathbf{\Psi}=&\big(\xi_{1s\uparrow},\cdots, \xi_{Ns\uparrow},\eta_{1s\uparrow},\cdots,\eta_{Ns\uparrow},\xi_{1p\uparrow},\cdots, \xi_{Np\uparrow},\eta_{1p\uparrow},\cdots,\eta_{Np\uparrow},\nonumber&\\
&\xi_{1s\downarrow},\cdots, \xi_{Ns\downarrow},\eta_{1s\downarrow},\cdots,\eta_{Ns\downarrow},\xi_{1p\downarrow},\cdots, \xi_{Np\downarrow},\eta_{1p\downarrow},\cdots,\eta_{Np\downarrow}\big)^T&
\end{align}
\subsubsection{I-matrix}
The $I$-matrix can be represented in a block structure within spin space.
\begin{align}
  \mathds{I} = \left[\begin{array}{ c | c }
    \mathds{I}^{\uparrow\uparrow} & \mathds{I}^{\uparrow\downarrow} \\
    \hline
    \mathds{I}^{\downarrow\uparrow} & \mathds{I}^{\downarrow\downarrow}
  \end{array}\right]
\end{align}

The components of the $\mathds{I}^{\sigma\sigma'}$ matrices are given by
\begin{align}
&\mathds{I}^{\sigma\sigma}_{i,j}=\delta_{ij}\left(1-n^{\overline{\sigma}}_s(i)\right)& \\
&\mathds{I}^{\sigma\sigma}_{i+N,j+N}=\delta_{ij}\left(n^{\overline{\sigma}}_s(i)\right)& \\
&\mathds{I}^{\sigma\sigma}_{i+N,j+N}=\delta_{ij}\left(1-n^{\overline{\sigma}}_p(i)\right)& \\
&\mathds{I}^{\sigma\sigma}_{i+N,j+N}=\delta_{ij}\left(n^{\overline{\sigma}}_p(i)\right)& \\
\end{align}
and
\begin{align}
&\mathds{I}^{\sigma\overline{\sigma}}_{i,i}=\delta_{ij}\left(S^-_s(i)\right)& \\
&\mathds{I}^{\sigma\overline{\sigma}}_{i+N,i+N}=\delta_{ij}\left(-S^-_s(i)\right)& \\
&\mathds{I}^{\sigma\overline{\sigma}}_{i+N,i+N}=\delta_{ij}\left(S^-_p(i)\right)& \\
&\mathds{I}^{\sigma\overline{\sigma}}_{i+N,i+N}=\delta_{ij}\left(-S^-_p(i)\right)& \\
\end{align}

\subsubsection{M-matrix}

The $M$-matrix can also be represented in a block structure within the spin space.
\begin{align}
  \mathds{M} = \left[\begin{array}{ c | c }
    \mathds{M}^{\uparrow\uparrow} & \mathds{M}^{\uparrow\downarrow} \\
    \hline
    \mathds{M}^{\downarrow\uparrow} & \mathds{M}^{\downarrow\downarrow}
  \end{array}\right]
\end{align}
Each $\mathds{M}^{\sigma\sigma'}$ matrix consists of local contributions from the atomic Hamiltonian as well as contributions from the hopping terms.
We can decompose each matrix into two parts:  $\mathds{M}^{0,\sigma\sigma'}$, which includes the local contributions and the spin-diagonal hopping terms, and $\mathds{M}^{\alpha,\sigma\sigma'}$ , which accounts for the contributions from the spin-orbit coupling term. Consequently, $\mathds{M}^{\sigma\sigma'}$ can be written as $\mathds{M}^{\sigma\sigma'}=\mathds{M}^{0,\sigma\sigma'}+\mathds{M}^{\alpha,\sigma\sigma'}$. The intra- and inter-orbital components are given by
\begin{align}
    &\mathds{M}^{0,\sigma\sigma}_{i,j} = -\delta_{ij} \left[(\mu+M) \left(1-n^{\overline{\sigma}}_s(i)\right) + te^{0,\sigma}_{s}(i) - te^{0,\sigma}_{sp}(i)\right] 
    - t \left[1-\left(n^{\overline{\sigma}}_s(i) + n^{\overline{\sigma}}_s(j)\right) + p^{0,\sigma}_{ss}(i,j)\right] \label{eq:M11} \\
    &\mathds{M}^{0,\sigma\sigma}_{i,j+N} = \delta_{ij}\left[te^{0,\sigma}_s(i) - te^{0,\sigma}_{sp}(i)\right] 
    - t \left(n^{\overline{\sigma}}_s(j) - p^{0,\sigma}_{ss}(i,j)\right) \label{eq:M12} \\
    &\mathds{M}^{0,\sigma\sigma}_{i+N,j} = \delta_{ij}\left[te^{0,\sigma}_s(i) - te^{0,\sigma}_{sp}(i)\right] 
    - t \left(n^{\overline{\sigma}}_s(i) - p^{0,\sigma}_{ss}(i,j)\right) \label{eq:M21} \\
    &\mathds{M}^{0,\sigma\sigma}_{i+N,j+N} = -\delta_{ij} \left[\left(\mu+M - U_s\right) n^{\overline{\sigma}}_s(i) + te^{0,\sigma}_s(i) - te^{0,\sigma}_{sp}(i)\right] 
    - t p^{0,\sigma}_{ss}(i,j) \label{eq:M22} \\
    &\mathds{M}^{0,\sigma\sigma}_{i+2N,j+2N} = -\delta_{ij} \left[(\mu-M) \left(1-n^{\overline{\sigma}}_p(i)\right) - te^{0,\sigma}_p(i) + te^{0,\sigma}_{ps}(i)\right] 
    + t \left(1 - \left[n^{\overline{\sigma}}_p(i) + n^{\overline{\sigma}}_p(j)\right] + p^{0,\sigma}_{pp}(i,j)\right) \label{eq:M33} \\
    &\mathds{M}^{0,\sigma\sigma}_{i+2N,j+3N} = \delta_{ij} \left[-te^{0,\sigma}_p(i) + te^{0,\sigma}_{ps}(i)\right] 
    + t \left(n^{\overline{\sigma}}_p(j) - p^{0,\sigma}_{pp}(i,j)\right) \label{eq:M34} \\
    &\mathds{M}^{0,\sigma\sigma}_{i+3N,j+2N} = \delta_{ij} \left[-te^{0,\sigma}_p(i) + te^{0,\sigma}_{ps}(i)\right] 
    + t \left(n^{\overline{\sigma}}_p(i) - p^{0,\sigma}_{pp}(i,j)\right) \label{eq:M43} \\
    &\mathds{M}^{0,\sigma\sigma}_{i+3N,j+3N} = -\delta_{ij} \left[\left(\mu-M - U_p\right) n^{\overline{\sigma}}_p(i) - te^{0,\sigma}_p(i) + te^{0,\sigma}_{ps}(i)\right] 
    + t p^{0,\sigma}_{pp}(i,j) \label{eq:M44}
\end{align}

\begin{align}
    &\mathds{M}^{\alpha,\sigma\sigma}_{i,j} = \delta_{ij}\alpha^{SOC} e^{\alpha,\sigma}_{s}(i) 
    +\alpha^{SOC} \left[S^-_s(i) + S^+_s(j) - p^{\alpha,\sigma}_{ss}(i,j)\right] \label{eq:M11} \\
    &\mathds{M}^{\alpha,\sigma\sigma}_{i,j+N} = -\delta_{ij}\alpha^{SOC}e^{\alpha,\sigma}_s(i) 
    +\alpha^{SOC} \left(-S^+_s(j) + p^{\alpha,\sigma}_{ss}(i,j)\right) \label{eq:M12} \\
    &\mathds{M}^{\alpha,\sigma\sigma}_{i+N,j} = -\delta_{ij}\alpha^{SOC}e^{\alpha,\sigma}_s(i)
    +\alpha^{SOC} \left(-S^-_s(i) + p^{\alpha,\sigma}_{ss}(i,j)\right) \label{eq:M21} \\
    &\mathds{M}^{0,\sigma\sigma}_{i+N,j+N} = \delta_{ij}\alpha^{SOC} e^{\alpha,\sigma}_s(i) 
    - \alpha^{SOC} p^{\alpha,\sigma}_{ss}(i,j) \label{eq:M22} \\
    &\mathds{M}^{\alpha,\sigma\sigma}_{i+2N,j+2N} = \delta_{ij} \alpha^{SOC} e^{\alpha,\sigma}_p(i)
    + \alpha^{SOC} \left[ S^-_p(i) + S^+_p(j) - p^{\alpha,\sigma}_{pp}(i,j)\right] \label{eq:M33} \\
    &\mathds{M}^{\alpha,\sigma\sigma}_{i+2N,j+3N} = -\delta_{ij} \alpha^{SOC}e^{\alpha,\sigma}_p(i) 
    + \alpha^{SOC} \left(-S^+_p(j) + p^{\alpha,\sigma}_{pp}(i,j)\right) \label{eq:M34} \\
    &\mathds{M}^{\alpha,\sigma\sigma}_{i+3N,j+2N} = -\delta_{ij} \alpha^{SOC}e^{\alpha,\sigma}_p(i)
    + \alpha^{SOC} \left(-S^-_p(i) + p^{\alpha\sigma}_{pp}(i,j)\right) \label{eq:M43} \\
    &\mathds{M}^{\alpha,\sigma\sigma}_{i+3N,j+3N} = \delta_{ij}\alpha^{SOC} e^{\alpha,\sigma}_p(i) 
    -\alpha^{SOC} p^{\alpha,\sigma}_{pp}(i,j) \label{eq:M44}
\end{align}

\begin{align}
    &\mathds{M}^{0,\sigma\sigma}_{i,i+\delta_y+2N} = -a(i,i+\delta_y) t\left(1-\left[n^{\overline{\sigma}}_s(i)+n^{\overline{\sigma}}_p(i+\delta_y)\right]+p^{0,\sigma}_{sp}(i,i+\delta_y)\right) \label{eq:M13y} \\
    &\mathds{M}^{0,\sigma\sigma}_{i+2N,i+\delta_y} = a(i,i+\delta_y) t\left(1-\left[n^{\overline{\sigma}}_s(i+\delta_y)+n^{\overline{\sigma}}_p(i)\right]+p^{0,\sigma}_{sp}(i,i+\delta_y)\right) \label{eq:M31y} \\
    &\mathds{M}^{0,\sigma\sigma}_{i,i+\delta_y+3N} = -a(i,i+\delta_y) t\left(n^{\overline{\sigma}}_p(i+\delta_y)-p^{0,\sigma}_{sp}(i,i+\delta_y)\right) \label{eq:M14y} \\
    &\mathds{M}^{0,\sigma\sigma}_{i+3N,i+\delta_y+N} = a(i,i+\delta_y) t\left(n^{\overline{\sigma}}_p(i)-p^{0,\sigma}_{sp}(i,i+\delta_y)\right) \label{eq:M41y} \\
    &\mathds{M}^{0,\sigma\sigma}_{i+N,i+\delta_y+2N} = -a(i,i+\delta_y) t\left(n^{\overline{\sigma}}_s(i)-p^{0,\sigma}_{sp}(i,i+\delta_y)\right) \label{eq:M23y} \\
    &\mathds{M}^{0,\sigma\sigma}_{i+2N,i+\delta_y+N} = a(i,i+\delta_y) t\left(n^{\overline{\sigma}}_s(i+\delta_y)-p^{0,\sigma}_{sp}(i,i+\delta_y)\right) \label{eq:M32y} \\
    &\mathds{M}^{0,\sigma\sigma}_{i+N,i+\delta_y+3N} = -a(i,i+\delta_y) tp^{0,\sigma}_{sp}(i,i+\delta_y) \label{eq:M24y} \\
    &\mathds{M}^{0,\sigma\sigma}_{i+3N,i+\delta_y+N} = a(i,i+\delta_y) tp^{0,\sigma}_{sp}(i,i+\delta_y) \label{eq:M42y}
\end{align}

\begin{align}
    &\mathds{M}^{0,\sigma\overline{\sigma}}_{i,j} = -\delta_{ij} \left[(\mu+M) S^-_s(i) + te^{0,\sigma\overline{\sigma}}_{s}(i) - te^{0,\sigma\overline{\sigma}}_{sp}(i)\right] 
    - t \left[-\left(S^-_s(i) + S^-_s(j)\right) + p^{0,\sigma\overline{\sigma}}_{ss}(i,j)\right] \label{eq:M11} \\
    &\mathds{M}^{0,\sigma\overline{\sigma}}_{i,j+N} = \delta_{ij}\left[te^{0,\sigma\overline{\sigma}}_s(i) - te^{0,\sigma\overline{\sigma}}_{sp}(i)\right] 
    - t \left(S^-_s(j) - p^{0,\sigma\overline{\sigma}}_{ss}(i,j)\right) \label{eq:M12} \\
    &\mathds{M}^{0,\sigma\overline{\sigma}}_{i+N,j} = \delta_{ij}\left[te^{0,\sigma\overline{\sigma}}_s(i) - te^{0,\sigma\overline{\sigma}}_{sp}(i)\right] 
    - t \left(S^-_s(i) - p^{0,\sigma\overline{\sigma}}_{ss}(i,j)\right) \label{eq:M21} \\
    &\mathds{M}^{0,\sigma\overline{\sigma}}_{i+N,j+N} = \delta_{ij} \left[\left(\mu+M - U_s\right) S^-_s(i) + te^{0,\sigma\overline{\sigma}}_s(i) - te^{0,\sigma\overline{\sigma}}_{sp}(i)\right] 
    - t p^{0,\sigma\overline{\sigma}}_{ss}(i,j) \label{eq:M22} \\
    &\mathds{M}^{0,\sigma\overline{\sigma}}_{i+2N,j+2N} = \delta_{ij} \left[(\mu-M) S^-_p(i) - te^{0,\sigma\overline{\sigma}}_p(i) + te^{0,\sigma\overline{\sigma}}_{ps}(i)\right] 
    + t \left( - \left[S^-_p(i) + S^-_p(j)\right] + p^{0,\sigma\overline{\sigma}}_{pp}(i,j)\right) \label{eq:M33} \\
    &\mathds{M}^{0,\sigma\overline{\sigma}}_{i+2N,j+3N} = \delta_{ij} \left[-te^{0,\sigma\overline{\sigma}}_p(i) + te^{0,\sigma\overline{\sigma}}_{ps}(i)\right] 
    + t \left(S^-_p(j) - p^{0,\sigma\overline{\sigma}}_{pp}(i,j)\right) \label{eq:M34} \\
    &\mathds{M}^{0,\sigma\overline{\sigma}}_{i+3N,j+2N} = \delta_{ij} \left[-te^{0,\sigma\overline{\sigma}}_p(i) + te^{0,\sigma\overline{\sigma}}_{ps}(i)\right] 
    + t \left(S^-_p(i) - p^{0,\sigma\overline{\sigma}}_{pp}(i,j)\right) \label{eq:M43} \\
    &\mathds{M}^{0,\sigma\overline{\sigma}}_{i+3N,j+3N} = \delta_{ij} \left[\left(\mu-M - U_p\right) S^-_p(i) - te^{0,\sigma\overline{\sigma}}_p(i) + te^{0,\sigma\overline{\sigma}}_{ps}(i)\right] 
    + t p^{0,\sigma\overline{\sigma}}_{pp}(i,j) \label{eq:M44}
\end{align}

\begin{align}
    &\mathds{M}^{\alpha,\sigma\overline{\sigma}}_{i,j} = \delta_{ij} \alpha^{SOC} e^{\alpha,\sigma\overline{\sigma}}_{s}(i) 
    + \alpha^{SOC} \left(1-\left[n^{\overline{\sigma}}_s(i) + n^{\sigma}_s(j)\right] + p^{\alpha,\sigma\overline{\sigma}}_{ss}(i,j)\right) \label{eq:M11} \\
    &\mathds{M}^{\alpha,\sigma\overline{\sigma}}_{i,j+N} = -\delta_{ij} \alpha^{SOC}e^{\alpha,\sigma\overline{\sigma}}_s(i) 
    + \alpha^{SOC} \left(n^{\overline{\sigma}}_s(j) - p^{\alpha,\sigma\overline{\sigma}}_{ss}(i,j)\right) \label{eq:M12} \\
    &\mathds{M}^{\alpha,\sigma\overline{\sigma}}_{i+N,j} = -\delta_{ij} \alpha^{SOC}e^{\alpha,\sigma\overline{\sigma}}_s(i)
    + \alpha^{SOC} \left(n^{\sigma}_s(i) - p^{\alpha,\sigma\overline{\sigma}}_{ss}(i,j)\right) \label{eq:M21} \\
    &\mathds{M}^{\alpha,\sigma\overline{\sigma}}_{i+N,j+N} = \delta_{ij} \alpha^{SOC} e^{\alpha,\sigma\overline{\sigma}}_s(i) 
    + \alpha^{SOC} p^{\alpha,\sigma\overline{\sigma}}_{ss}(i,j) \label{eq:M22} \\
    &\mathds{M}^{\alpha,\sigma\overline{\sigma}}_{i+2N,j+2N} = \delta_{ij}\alpha^{SOC}e^{\alpha,\sigma\overline{\sigma}}_p(i)
    + \alpha^{SOC} \left(1 - \left[n^{\sigma}_p(i) + n^{\overline{\sigma}}_p(j)\right] + p^{\alpha,\sigma\overline{\sigma}}_{pp}(i,j)\right) \label{eq:M33} \\
    &\mathds{M}^{\alpha,\sigma\overline{\sigma}}_{i+2N,j+3N} = -\delta_{ij} \alpha^{SOC}e^{\alpha,\sigma\overline{\sigma}}_p(i)
    + \alpha^{SOC} \left(n^{\overline{\sigma}}_p(j) - p^{\alpha,\sigma\overline{\sigma}}_{pp}(i,j)\right) \label{eq:M34} \\
    &\mathds{M}^{\alpha,\sigma\overline{\sigma}}_{i+3N,j+2N} = -\delta_{ij} \alpha^{SOC}e^{\alpha,\sigma\overline{\sigma}}_p(i) 
    + \alpha^{SOC} \left(n^{\sigma}_p(i) - p^{\alpha,\sigma\overline{\sigma}}_{pp}(i,j)\right) \label{eq:M43} \\
    &\mathds{M}^{\alpha,\sigma\overline{\sigma}}_{i+3N,j+3N} = \delta_{ij} \alpha^{SOC}e^{\alpha,\sigma\overline{\sigma}}_p(i) 
    + \alpha^{SOC} p^{\alpha,\sigma\overline{\sigma}}_{pp}(i,j) \label{eq:M44}
\end{align}

\begin{align}
 &\mathds{M}^{0,\sigma\overline{\sigma}}_{i,i+\delta_y+2N} = -a(i,i+\delta_y) t\left(-\left[S^-_s(i)+S^-_p(i+\delta_y)\right]+p^{0,\sigma\overline{\sigma}}_{sp}(i,i+\delta_y)\right) \label{eq:M13y} \\   
 &\mathds{M}^{0,\sigma\overline{\sigma}}_{i+2N,i+\delta_y} = a(i,i+\delta_y) t\left(-\left[S^+_s(i+\delta_y)+S^+_p(i)\right]+p^{0,\sigma\overline{\sigma}}_{sp}(i,i+\delta_y)\right) \label{eq:M31y} \\
 &\mathds{M}^{0,\sigma\overline{\sigma}}_{i,i+\delta_y+3N} = -a(i,i+\delta_y) t\left(S^-_p(i+\delta_y)-p^{0,\sigma\overline{\sigma}}_{sp}(i,i+\delta_y)\right) \label{eq:M14y} \\
 &\mathds{M}^{0,\sigma\overline{\sigma}}_{i+3N,i+\delta_y+N} = a(i,i+\delta_y) t\left(S^+_p(i)-p^{0,\sigma\overline{\sigma}}_{sp}(i,i+\delta_y)\right) \label{eq:M41y} \\
 &\mathds{M}^{0,\sigma\overline{\sigma}}_{i+N,i+\delta_y+2N} = -a(i,i+\delta_y) t\left(S^-_s(i)-p^{0,\sigma\overline{\sigma}}_{sp}(i,i+\delta_y)\right) \label{eq:M23y} \\
 &\mathds{M}^{0,\sigma\overline{\sigma}}_{i+2N,i+\delta_y+N} = a(i,i+\delta_y) t\left(S^+_s(i+\delta_y)-p^{0,\sigma\overline{\sigma}}_{sp}(i,i+\delta_y)\right) \label{eq:M32y} \\
 &\mathds{M}^{0,\sigma\overline{\sigma}}_{i+N,i+\delta_y+3N} = -a(i,i+\delta_y) tp^{0,\sigma\overline{\sigma}}_{sp}(i,i+\delta_y) \label{eq:M24y} \\
 &\mathds{M}^{0,\sigma\overline{\sigma}}_{i+3N,i+\delta_y+N} = a(i,i+\delta_y) tp^{0,\sigma\overline{\sigma}}_{sp}(i,i+\delta_y) \label{eq:M42y}
\end{align}

\subsubsection{Self-consistent parameters\label{App:BHZ_SC_params}}

The $p$ parameters can be expressed as a sum of correlation functions involving bosonic operators. However, they cannot be directly computed from $\mathds{C}$, requiring a decoupling scheme to relate them to $\mathds{C}$. Here $\gamma,\tau$ are orbital indices
\begin{align}
&p^{0,\downarrow}_{\gamma\tau}(i,j) = \langle n^{\downarrow}_\gamma(i)n^{\downarrow}_\tau(j)\rangle + \langle S^-_\gamma(i) S^+_\tau(j)\rangle - \langle \Delta_\gamma(i) \Delta\tau^\dagger(j)\rangle &\\
&p^{0,\uparrow}_{\gamma\tau}(i,j) =  \langle n^{\uparrow}_\gamma(i)n^{\uparrow}_\tau(j)\rangle + \langle S^+_\gamma(i) S^-_\tau(j)\rangle - \langle \Delta_\gamma(i) \Delta_\tau^\dagger(j)\rangle  &\\
&p^{0,\uparrow\downarrow}_{\gamma\tau}(i,j) = \langle n^{\downarrow}_\gamma(i)S^-\tau(j) \rangle + \langle S^-_\gamma(i)n^{\uparrow}\tau(j) \rangle&\\
&p^{\alpha,\downarrow}_{\gamma\tau}(i,j)=\langle n^{\downarrow}_\gamma(i)S^+_\tau(j)\rangle+\langle S^-_\gamma(i)n^{\downarrow}_\tau(j)\rangle &\\
&p^{\alpha,\uparrow}_{\gamma\tau}(i,j)= \langle  S^+_\gamma(i)n^{\uparrow}_\tau(j)\rangle + \langle S^+_\gamma(i) n^{\downarrow}_\tau(j)\rangle&\\
&p^{\alpha,\uparrow\downarrow}_{\gamma\tau}(i,j)= \langle n^{\downarrow}_\gamma(i)n^{\uparrow}_\tau(j)\rangle + \langle S^-_\gamma(i) S^-_\tau(j)\rangle + \langle \Delta_\gamma(i) \Delta^\dagger_\tau(j)\rangle&
\end{align}
The expectation values of the density and spin-flip operators are single-particle quantities that can be computed using the on-site composite correlation matrix $\mathds{C}$.
\begin{align}
&n^\sigma_\gamma(i)=\langle (c_\gamma^{\sigma})^\dagger(i)c_\gamma^\sigma(i)\rangle&\\
&S_\gamma^-(i)=\langle (c_\gamma^\downarrow)^\dagger(i) c^\uparrow_\gamma(i)\rangle&
\end{align} 
The $e$ parameters are single-particle quantities that can be computed using the composite correlation matrix $\mathds{C}$.
\begin{align}
&e_{\tau }^{0,\sigma}(i)=\sum_{\langle i,j \rangle} \left(\langle c_{j \tau \overline{\sigma}}\xi^\dagger_{i \tau\overline{\sigma}}\rangle+\langle c^\dagger_{j \tau \overline{\sigma}}\eta_{i \tau \overline{\sigma}}\rangle\right)&\\
&e_{\tau }^{\alpha,\sigma}(i)=\sum_{\langle i,j \rangle} \left(\langle c_{j \tau \sigma}\xi^\dagger_{i \tau\overline{\sigma}}\rangle+\langle c^\dagger_{j \tau \sigma}\eta_{i \tau \overline{\sigma}}\rangle\right)&\\
&e_{\tau }^{0,\sigma\overline{\sigma}}(i)=\sum_{\langle i,j \rangle} \left(\langle c_{j \tau \sigma}\xi^\dagger_{i \tau\overline{\sigma}}\rangle+\langle c^\dagger_{j \tau \overline{\sigma}}\eta_{i \tau \sigma}\rangle\right)&\\
&e_{\tau }^{\alpha,\sigma\overline{\sigma}}(i)=\sum_{\langle i,j \rangle} \left(\langle c_{j \tau \sigma}\xi^\dagger_{i \tau\sigma}\rangle+\langle c^\dagger_{j \tau \overline{\sigma}}\eta_{i \tau \overline{\sigma}}\rangle\right)&
\end{align}

\begin{align}
&e^{0,\sigma}_{sp}(i)=\sum_{\langle i,j \rangle_{\hat{x}}}a(i,j)\left(\langle c_{jp\overline{\sigma}}\xi^\dagger_{is\overline{\sigma}}\rangle+\langle c^\dagger_{jp\overline{\sigma}}\eta_{is\overline{\sigma}}\rangle\right)+\sum_{\langle i,j \rangle_{\hat{y}},} a(i,j)\left(\langle c_{jp\overline{\sigma}}\xi^\dagger_{is\overline{\sigma}}\rangle-\langle c^\dagger_{jp\overline{\sigma}}\eta_{is\overline{\sigma}}\rangle\right)&\\
&e^{0,\sigma}_{ps}(i)=\sum_{\langle i,j \rangle_{\hat{x}}}a(i,j)\left(\langle c_{js\overline{\sigma}}\xi^\dagger_{ip\overline{\sigma}}\rangle+\langle c^\dagger_{js\overline{\sigma}}\eta_{ip\overline{\sigma}}\rangle\right)+\sum_{\langle i,j\rangle_{\hat{y}} }a(i,j)\left(\langle c_{js\overline{\sigma}}\xi^\dagger_{ip\overline{\sigma}}\rangle-\langle c^\dagger_{js\overline{\sigma}}\eta_{ip\overline{\sigma}}\rangle\right)&\\
&e^{\alpha,\sigma}_{sp}(i)=\sum_{\langle i,j \rangle_{\hat{x}}}a(i,j)\left(\langle c_{jp\sigma}\xi^\dagger_{is\overline{\sigma}}\rangle+\langle c^\dagger_{jp\sigma}\eta_{is\overline{\sigma}}\rangle\right)+\sum_{\langle i,j \rangle_{\hat{y}},} a(i,j)\left(\langle c_{jp\sigma}\xi^\dagger_{is\overline{\sigma}}\rangle-\langle c^\dagger_{jp\sigma}\eta_{is\overline{\sigma}}\rangle\right)&\\
&e^{\alpha,\sigma}_{ps}(i)=\sum_{\langle i,j \rangle_{\hat{x}}}a(i,j)\left(\langle c_{js\sigma}\xi^\dagger_{ip\overline{\sigma}}\rangle+\langle c^\dagger_{js\sigma}\eta_{ip\overline{\sigma}}\rangle\right)+\sum_{\langle i,j\rangle_{\hat{y}} }a(i,j)\left(\langle c_{js\sigma}\xi^\dagger_{ip\overline{\sigma}}\rangle-\langle c^\dagger_{js\sigma}\eta_{ip\overline{\sigma}}\rangle\right)&\\
\end{align}

\begin{align}
&e^{0,\sigma\overline{\sigma}}_{sp}(i)=\sum_{\langle i,j \rangle_{\hat{x}}}a(i,j)\left(\langle c_{jp\sigma}\xi^\dagger_{is\overline{\sigma}}\rangle+\langle c^\dagger_{jp\overline{\sigma}}\eta_{is\sigma}\rangle\right)+\sum_{\langle i,j \rangle_{\hat{y}},} a(i,j)\left(\langle c_{jp\sigma}\xi^\dagger_{is\overline{\sigma}}\rangle-\langle c^\dagger_{jp\overline{\sigma}}\eta_{is\sigma}\rangle\right)&\\
&e^{0,\sigma\overline{\sigma}}_{ps}(i)=\sum_{\langle i,j \rangle_{\hat{x}}}a(i,j)\left(\langle c_{js\sigma}\xi^\dagger_{ip\overline{\sigma}}\rangle+\langle c^\dagger_{js\overline{\sigma}}\eta_{ip\sigma}\rangle\right)+\sum_{\langle i,j\rangle_{\hat{y}} }a(i,j)\left(\langle c_{js\sigma}\xi^\dagger_{ip\overline{\sigma}}\rangle-\langle c^\dagger_{js\overline{\sigma}}\eta_{ip\sigma}\rangle\right)&\\
&e^{\alpha,\sigma\overline{\sigma}}_{sp}(i)=\sum_{\langle i,j \rangle_{\hat{x}}}a(i,j)\left(\langle c_{jp\sigma}\xi^\dagger_{is\sigma}\rangle+\langle c^\dagger_{jp\sigma}\eta_{is\overline{\sigma}}\rangle\right)+\sum_{\langle i,j \rangle_{\hat{y}},} a(i,j)\left(\langle c_{jp\sigma}\xi^\dagger_{is\sigma}\rangle-\langle c^\dagger_{jp\sigma}\eta_{is\overline{\sigma}}\rangle\right)&\\
&e^{\alpha,\sigma\overline{\sigma}}_{ps}(i)=\sum_{\langle i,j \rangle_{\hat{x}}}a(i,j)\left(\langle c_{js\sigma}\xi^\dagger_{ip\sigma}\rangle+\langle c^\dagger_{js\sigma}\eta_{ip\overline{\sigma}}\rangle\right)+\sum_{\langle i,j\rangle_{\hat{y}} }a(i,j)\left(\langle c_{js\sigma}\xi^\dagger_{ip\sigma}\rangle-\langle c^\dagger_{js\sigma}\eta_{ip\overline{\sigma}}\rangle\right)&\\
\end{align}



\subsubsection{Roth decoupling\label{App:Roth}}

We employ the Roth decoupling scheme~\cite{roth1969electron} to compute two-point correlation functions, including density-density, spin-spin, and pair-pair correlations. This approach is essential for determining the $p$-parameters, which cannot be directly expressed in terms of the composite correlation matrix $\mathds{C}$ without a decoupling scheme. This allows us to express $p$ in terms of both on-site and intersite correlations. The detailed formalism can be found in Refs.\cite{pangburn2024spontaneous,roth1969electron,banerjee2024charge}. Here, we present the expressions for the bosonic-bosonic correlation functions in the Chern-Hubbard model, with $\tau, \gamma$ orbital indices.

\begin{align}
\label{eq:Roth_DeltaDelta}&\langle\Delta_{i,\tau}\Delta^\dagger_{j,\gamma} \rangle=\dfrac{\rho^\Delta_{\tau\gamma}(i,j)}{1-\phi_{\tau\gamma}(i)}&\\
\label{eq:Roth_SplusSminus}&\langle S^-_{i,\tau}S^+_{j,\gamma} \rangle=-\dfrac{\rho^S_{\tau\gamma}(i,j)}{1+\phi_{\tau\gamma}(i)}&\\
\label{eq:Roth_nn}&\langle n_{i,\tau\sigma}n_{j,\beta\sigma} \rangle=\dfrac{\left(-1-\phi_{\tau\gamma}(i)+2\left(\rho_{0,\tau\gamma}(i,j)+\phi_{\tau\gamma}(i)(\rho_{0,\tau\gamma}(i,j))+\rho^{n_\sigma n_\sigma} _{\tau\gamma}(i,j)\right)\right)}{2 \left(\phi_{\tau\gamma}(i)\right)^2-2}&
\end{align}

To simplify notation, we introduce the following on-site and nearest-neighbor correlation functions:
\begin{align}
&\mathbf{\Psi}(i)=\left(\xi_s(i),\eta_s(i),\xi_p(i),\eta_p(i)\right)^T&\\
&C_0^{\alpha\beta}(i)=\langle\mathbf{\Psi}_\alpha(i)\mathbf{\Psi}^\dagger_\beta(i)\rangle&\\
&C_1^{\alpha\beta}(i,j)=\langle\mathbf{\Psi}_\alpha(i)\mathbf{\Psi}^\dagger_\beta(j)\rangle \quad i,j \quad \text{nearest neighbour}&
\end{align}

Thus we have
\begin{align}
&\phi_{ss}(i)=\phi_{sp}^\Delta(i)=\dfrac{2}{2-n_s(i)}\left(C_0^{11}(i)+C_0^{21}(i)\right)-\dfrac{2}{n_s(i)}\left(C_0^{12}(i)+C_0^{22}(i)\right)&\\
&\phi_{pp}(i)=\phi_{ps}^\Delta(i)=\dfrac{2}{2-n_p(i)}\left(C_0^{33}(i)+C_0^{43}(i)\right)-\dfrac{2}{n_p(i)}\left(C_0^{34}(i)+C_0^{44}(i)\right)&\\
&\rho_{0,ss}(i)=\dfrac{2}{2-n_s(i)}\dfrac{n_s}{2}\left(C_0^{11}(i)+C_0^{21}(i)\right)&\\
&\rho_{0,pp}(i)=\dfrac{2}{2-n_p(i)}\dfrac{n_p(i)}{2}\left(C_0^{33}(i)+C_0^{43}(i)\right)&\\
&\rho_{0,sp}(i)=\dfrac{2}{2-n_s(i)}\dfrac{n_p(i)}{2}\left(C_0^{11}(i)+C_0^{21}(i)\right)&\\
&\rho_{0,ps}(i)=\dfrac{2}{2-n_p(i)}\dfrac{n_s(i)}{2}\left(C_0^{33}(i)+C_0^{43}(i)\right)&
\end{align}

\begin{align}
&\rho_{ss}^\Delta(i,j)=\dfrac{2}{2-n_s(i)}\left(C_1^{11}+C_1^{21}\right)\left(C_1^{21}+C_1^{22}\right)+\dfrac{2}{n_s(i)}\left(C_1^{11}+C_1^{12}\right)\left(C_1^{12}+C_1^{22}\right)&\\
&\rho_{pp}^\Delta(i,j)=\dfrac{2}{2-n_p(i)}\left(C_1^{33}+C_1^{43}\right)\left(C_1^{43}+C_1^{44}\right)+\dfrac{2}{n_p(i)}\left(C_1^{33}+C_1^{34}\right)\left(C_1^{34}+C_1^{44}\right)&\\
&\rho_{sp}^\Delta(i,j)=\dfrac{2}{2-n_p(i)}\left(C_1^{13}+C_1^{23}\right)\left(C_1^{41}+C_1^{42}\right)+\dfrac{2}{n_p(i)}\left(C_1^{31}+C_1^{32}\right)\left(C_1^{14}+C_1^{24}\right)&\\
&\rho_{ps}^\Delta(i,j)=\dfrac{2}{2-n_s(i)}\left(C_1^{31}+C_1^{41}\right)\left(C_1^{23}+C_1^{24}\right)+\dfrac{2}{n_s(i)}\left(C_1^{13}+C_1^{14}\right)\left(C_1^{32}+C_1^{42}\right)&
\end{align}
\begin{align}
&\rho_{ss}^S(i,j)=\dfrac{2}{2-n_s(i)}\left(C_1^{11}+C_1^{21}\right)\left(C_1^{11}+C_1^{12}\right)+\dfrac{2}{n_s(i)}\left(C_1^{22}+C_1^{12}\right)\left(C_1^{21}+C_1^{22}\right)&\\
&\rho_{pp}^S(i,j)=\dfrac{2}{2-n_p(i)}\left(C_1^{33}+C_1^{43}\right)\left(C_1^{33}+C_1^{34}\right)+\dfrac{2}{n_p(i)}\left(C_1^{44}+C_1^{34}\right)\left(C_1^{43}+C_1^{44}\right)&\\
&\rho_{sp}^S(i,j)=\dfrac{2}{2-n_p(i)}\left(C_{1}^{23}+C_1^{13}\right)\left(C_1^{31}+C_1^{32}\right)+\dfrac{2}{n_p(i)}\left(C_1^{14}+C_1^{24}\right)\left(C_1^{41}+C_1^{42}\right)&\\
&\rho_{ps}^S(i,j)=\dfrac{2}{2-n_s(i)}\left(C_1^{41}+C_1^{31}\right)\left(C_1^{13}+C_1^{14}\right)+\dfrac{2}{n_s(i)}\left(C_1^{32}+C_1^{42}\right)\left(C_1^{23}+C_1^{24}\right)&
\end{align}
\begin{align}
&\rho_{ss}^{n_\sigma n_\sigma}(i,j)=\dfrac{2}{2-n_s(i)}\left(C_1^{11}+C_1^{21}\right)\left(C_1^{11}+C_1^{12}\right)+\dfrac{2}{n_s(i)}\left(C_1^{22}+C_1^{12}\right)\left(C_1^{21}+C_1^{22}\right)&\\
&\rho_{pp}^{n_\sigma n_\sigma}(i,j)=\dfrac{2}{2-n_p(i)}\left(C_1^{33}+C_1^{43}\right)\left(C_1^{33}+C_1^{34}\right)+\dfrac{2}{n_p(i)}\left(C_1^{44}+C_1^{34}\right)\left(C_1^{43}+C_1^{44}\right)&\\
&\rho_{sp}^{n_\sigma n_\sigma}(i,j)=\dfrac{2}{2-n_p(i)}\left(C_1^{13}+C_1^{23}\right)\left(C_1^{31}+C_1^{32}\right)+\dfrac{2}{n_p(i)}\left(C_1^{14}+C_{1}^{24}\right)\left(C_{1}^{41}+C_1^{42}\right)&\\
&\rho_{ps}^{n_\sigma n_\sigma}(i,j)=\dfrac{2}{2-n_s(i)}\left(C_1^{31}+C_1^{41}\right)\left(C_1^{13}+C_1^{14}\right)+\dfrac{2}{n_s(i)}\left(C_1^{32}+C_1^{42}\right)\left(C_1^{23}+C_1^{24}\right)&
\end{align}
Where all the $C^{ab}_1=C^{ab}_1(i,j)$.\\

The Roth decoupling for the BHZ-Hubbard model without the paramagnetic approximation is significantly more challenging. However, since our focus in this work is not on determining the self-consistent ground state but rather on exploring topological properties, we adopt the same decoupling scheme for the BHZ-Hubbard model as for the Chern-Hubbard model. Additionally, we assume that the $p$-terms, which cannot be directly computed using the Chern-Hubbard-decoupling, vanish.


\section{Choosen set of parameters\label{App:Params}}

In this appendix, we present the composite operator parameters used to obtain the results shown in the main text for both the Chern-Hubbard model and the BHZ-Hubbard model.

\begin{table}[h!]
\centering
\begin{tabular}{|c|c|c|c|c|c|c|c|c|c|c|c|c|c|}
\hline
Phase & $\mu$ & $n_s$ & $n_p$ & $e_s$ & $e_p$ & $p_s$ & $p_p$ & $e_{sp}$ & $e_{ps}$ & $p_{sp}$ & $M$ & $U_s$ & $U_p$ \\ 
\hline 
MI & $6.$ & $1$ & $1$ & $-1.2$ & $-1.2$ & $0.2$ & $0.2 $& $-0.5$ & $-0.5$ & $0.2$ & $0.$ & $12$ & $12$ \\ 
\hline
TMBI & $6.$ & $1$ & $1$ & $-4.8$ & $-4.8$ & $0.2$ & $0.2 $& $-0.5$ & $-0.5$ & $0.2$ & $0.$ & $12$ & $12$ \\ 
\hline
TMZ & $6.$ & $1$ & $1$ & $-0.4$ & $-0.4$ & $0.2$ & $0.2 $& $-0.5$ & $-0.5$ & $0.2$ & $0.$ &$12$ & $12$ \\ 
\hline
\end{tabular}
\caption{The set of parameters used in the Chern-Hubbard model for the results presented in the main text. The first row corresponds to the trivial Mott insulator (MI) phase, the second row to the topological Mott band insulator (TMBI) phase, and the third row to the topological Mott zeros (TMZ) phase.}
\label{Table:Params_Chern}
\end{table}

\begin{table}[h!]
\centering
\begin{tabular}{|c|c|c|c|c|c|c|c|c|c|c|c|c|c|c|c|c|c|c|c|c|c|c|}
\hline
Phase & $\mu$ & $n_s$ & $n_p$ & $e^{0,\sigma}_s$ & $e^{0,\sigma}_p$ & $p^{0,\sigma}_s$ & $p^{0,\sigma}_p$ & $e^{0,\sigma}_{sp}$ & $e^{0,\sigma}_{ps}$ & $p^{0,\sigma}_{sp}$ & $e_s^{0,\sigma\overline{\sigma}}$ & $e_p^{0,\sigma\overline{\sigma}}$ & $e^{0,\sigma\overline{\sigma}}_{sp}$ & $e^{0,\sigma\overline{\sigma}}_{ps}$ &$p_s^{0,\sigma\overline{\sigma}}$ & $p_p^{0,\sigma\overline{\sigma}}$ & $p_{sp}^{0,\sigma\overline{\sigma}}$ &$M$ & $U_s$ & $U_p$ & $\alpha^{SOC}$ & $\Delta_\tau$\\ 
\hline 
TSMBI & $6.$ & $1$ & $1$ & $-1.2$ & $-1.2$ & $0.2$ & $0.2 $& $-0.5$ & $-0.5$ & $0.2$ & $0.$ & $0.$ &$-0.05$ & $-0.05$& $0$ & $0.$ & $0.$ &$0.$ & $12$ & $12$ & $0.1$ &  $0.$\\ 
\hline
TSMZ & $6.$ & $1$ & $1$ & $-0.4$ & $-0.4$ & $0.2$ & $0.2 $& $-0.5$ & $-0.5$ & $0.2$ & $0$& $0.$ & $-0.01$& $-0.01$& $0.$& $0.$& $0.$& $0.$ & $12$ & $12$ & $0.02$ &  $0.$\\ 
\hline
\end{tabular}
\caption{The set of parameters, excluding spin-orbit-related terms, used in the BHZ-Hubbard model for the results presented in the main text. The first row corresponds to the topological spin Mott band insulator (TSMBI) phase, while the second row corresponds to  the topological spin Mott zeros (TSMZ) phase.}
\label{Table:Params_BHZ_1}
\end{table}

\begin{table}[h!]
\centering
\begin{tabular}{|c|c|c|c|c|c|c|c|c|c|c|c|c|c|c|c|c|c|c|c|c|c|}
\hline
Phase & $\mu$ & $n_s$ & $n_p$ & $e^{\alpha,\sigma}_s$ & $e^{\alpha,\sigma}_p$ & $p^{\alpha,\sigma}_s$ & $p^{\alpha,\sigma}_p$ & $e^{\alpha,\sigma}_{sp}$ & $e^{\alpha,\sigma}_{ps}$ & $p^{\alpha,\sigma}_{sp}$ & $e_s^{\alpha,\sigma\overline{\sigma}}$ & $e_p^{\alpha,\sigma\overline{\sigma}}$ & $e^{\alpha,\sigma\overline{\sigma}}_{sp}$ & $e^{\alpha,\sigma\overline{\sigma}}_{ps}$ &$p_s^{\alpha,\sigma\overline{\sigma}}$ & $p_p^{\alpha,\sigma\overline{\sigma}}$  &$M$ & $U_s$ & $U_p$ & $\alpha^{SOC}$ & $\Delta_\tau$\\ 
\hline 
TSMBI & $6.$ & $1$ & $1$ & $0.$ & $0.$ & $0.$ & $0. $& $0.$ & $0.$ & $0.$ & $0.$ & $0$ &$0.$ & $0.$& $0.$& $0.$& $0.$ & $12$ & $12$ & $0.1$ &  $0.$\\ 
\hline
TSMZ & $6.$ & $1$ & $1$ & $ 0.$ & $ 0.$ & $0.$ & $0. $& $0.$ & $ 0.$ & $0.$ & $0$& $0$ & $0.$& $0.$& $0.$& $0.$& $0.$ & $12$ & $12$ & $0.02$ &  $0.$\\ 
\hline
\end{tabular}
\caption{The set of spin-orbit related parameters used in the BHZ-Hubbard model for the results presented in the main text. The first row corresponds to the topological spin Mott band insulator (TSMBI) phase, while the second row corresponds to the topological spin Mott zeros (TSMZ) phase.}
\label{Table:Params_BHZ_2}
\end{table}

\end{onecolumngrid}

\end{document}